\documentclass[12pt]{article}

\usepackage{amsmath,amsfonts,epsfig,color,latexsym}

\newcommand{\be}{\begin{equation}}
\newcommand{\ee}{\end{equation}}
\newcommand{\bea}{\begin{eqnarray}}
\newcommand{\eea}{\end{eqnarray}}

\let\newsection=\section
\renewcommand{\section}{\setcounter{equation}{0}\newsection}

\begin{document}

\begin{flushright}
hep-th/0409099\\
BROWN-HET-1421
\end{flushright}
\vskip.5in

\begin{center}

{\LARGE\bf Planckian scattering effects and black hole production in 
low $M_{Pl}$ scenarios }
\vskip 1in
\centerline{\Large Kyungsik Kang and Horatiu Nastase}
\vskip .5in

\end{center}
\centerline{\large Brown University}
\centerline{\large Providence, RI, 02912, USA}

\vskip 1in

\begin{abstract}

{\large We reanalyze the question of black hole creation in high energy 
scattering via shockwave collisions. We find that string corrections 
tend to increase the scattering cross-section. We analyze corrections 
in a more physical setting, of Randall-Sundrum type and of higher 
dimensionality. We also analyze the scattering inside AdS backgrounds. 
}

\end{abstract}

\newpage

\section{Introduction}

The problem of black hole creation in high energy scattering is one of 
significant importance, for two possible reasons. One is that one can have 
a low gravitational scale, as in the large extra dimensions 
\cite{add,aadd} or Randall-Sundrum \cite{rs, rs2}
scenarios. Thus the possibility of black hole creation at accelerators has 
been explored at length in a number of papers (e.g. 
\cite{gidbh,dl,bf,de,vol1,vol2,eg,jt,emparan}).

Another reason, is that 
via gauge-gravity dualities, the high energy scattering in a gravity theory 
can be related to high energy scattering in QCD, or gauge 
theories in general \cite{ps,gid,gkp}. Simply put, high energy scattering 
in QCD can be described in terms of a conformal field theory with a cutoff, 
and that is dual to a two brane Randall-Sundrum scenario. 
But then black hole creation that happens in high energy gravity scattering
has to have some implications for the QCD side. In fact, in \cite{gid} 
it was argued that black hole creation, when the black hole size is 
comparable to the size of the gravity dual= AdS slice, is responsible for 
the much sought-for Froissart behaviour (saturation of the unitarity 
bound). We will revisit these questions in a future paper \cite{inprog}, 
but we will still set up some of the calculations needed for that case 
in here.

In particular, we will analyze the case of high energy scattering with 
black hole formation inside AdS space. 

We will focus instead on the actual black hole creation at high energy 
$s\sim M_{Pl}^2$ with the idea of applying it to 
theories with a low fundamental scale. 

Giddings and Thomas \cite{gidbh} and a number 
of other people \cite{dl,de,vol1,vol2,jt} (see also earlier work in 
\cite{bf}) have proposed 
that the cross-section for black hole creation in flat space at high 
energy is just proportional to the geometric horizon area of a black 
hole of mass equal to the total center of mass energy, i.e.
\be
\sigma \simeq \pi r_H^2;\;\; r_H=2G\sqrt{s} \;\;(D=4)
\ee
There has been a considerable amount of debate over whether this assumption 
is correct (see, e.g. \cite{vol1,vol2,solo,krasnov,jt,emparan}). 

In an attempt to prove it, Eardley and Giddings \cite{eg} have treated the 
high energy collision according to a recipe proposed some time ago by 
't Hooft \cite{thooft}. The process is well described by the collision of 
two gravitational shockwaves of Aichelburg-Sexl type. Even though one cannot 
calculate precisely the metric in the future of the collision except 
perturbatively \cite{dp}, one can use a trick due to Penrose that just uses 
the properties of Einstein gravity to calculate a lower bound on the area 
of the horizon that will form in the collision. 

In D=4 \cite{eg} were able to extend Penrose's method to collision at nonzero 
impact parameter b of the two Aichelburg-Sexl waves, and prove that the 
cross section for black hole scattering is indeed of the order of magnitude 
of the geometric cross-section of the classical black hole. 

In this paper we will try to refine this calculation, and answer some of the 
criticisms addressed to the calculation and the geometric cross section 
result. One such criticism was that string 
corrections will significantly lower this result (see \cite{ry} for 
example) We will try to analyze 
string corrections explicity via two methods. 

There are two modifications 
of the Aichelburg-Sexl metric that were shown to reproduce string scattering
results (effective metrics). The one in \cite{acv93} analyzes specifically 
the scattering at impact parameter b, and gives an effective metric for 
large b ($>R_s$, the gravitational radius for black hole formation). It is 
therefore unsuited for our purposes, yet with some  approximations 
one can find that the head-on collison of two such waves (each having a 
parameter b) will have an increased horizon area of the formed black hole, 
with respect to the Aichelburg-Sexl case. The second modification \cite{ac}
corresponds to string-corrected 't Hooft scattering in an Aichelburg-Sexl 
metric. We will show that scattering of two modified shockwaves will 
again increase the horizon area of the formed black hole.

Another possible caveat to the calculation in \cite{eg} is that it was done 
in flat D=4. We will analyze the case of the more realistic Randall-Sundrum
scenario and find that we just get small corrections to the flat D=4 case. 
We will also offer a method of estimating the cross section in the arbitrary 
D case. 

The paper is organized as follows. In section 2 we will review the 
Aichelburg-Sexl wave and 't Hooft's scattering calculation, and generalize 
it to higher dimension. In section 3 we will review the analysis of \cite{eg}
and set it up for generalization to any shockwaves and any dimension. We 
will also analyze the collision of sourceless waves, which should describe 
graviton-graviton scattering, and present a puzzle. 
In section 4, we will analyze string corrections via the effective 
metrics in \cite{acv93} and \cite{ac}. In section 5 we analyze the 
case of Randall-Sundrum background and calculate corrections. 
In section 6 we will write down a solution for an Aichelburg-Sexl wave 
inside AdS and do a 't Hooft scattering analysis.

\section{The Aichelburg-Sexl wave and 't Hooft scattering at high energy}

't Hooft \cite{thooft}
has proposed that an (almost) massless particle at high energies 
$s\sim M_{Pl}^2\gg t$ behaves like a plane gravitational wave- a shockwave, 
and its only interactions are given by massless particles, 
with the gravitational interactions 
described by deflection in the gravitational shockwave corresponding to the 
massless particle. That shockwave
solution is due to Aichelburg and Sexl \cite{as}.

In this section we will review this procedure 
of gravitational interaction and generalize it to higher 
dimensions. 

The Aichelburg-Sexl solution is of the pp wave type.
A pp wave (plane fronted gravitational waves) has the general form 
in d dimensions
\be
ds^2 =-dx^+dx^-+(dx^+)^2H(x^+, x^i)+\sum_{i=1}^{d-2} (dx_i)^2
\ee
and has Ricci tensor 
\be
R_{++}=-1/2 \partial_i^2 H(x^+, x^i)
\ee 
and the rest are zero. Horowitz and Steif \cite{hs} showed that there are no 
quantum ($\alpha '$)
corrections to the (purely gravitational and NS-NS background) 
pp wave solutions, since all the gravitational invariants made from Ricci and 
Riemann tensors vanish on this solution. The inverse metric is given by 
\be
g^{\mu\nu}\partial_{\mu}\partial_{\nu}= -4\partial_+\partial_- -4H\partial_-^2
+\partial_i^2
\ee
and so for instance
\be
R^{(2)}\equiv R_{\mu\nu}R^{\mu\nu}=R_{\mu\nu}R_{\rho\sigma}g^{\mu\rho}
g^{\nu\sigma}
\ee
does not contain $(R_{++})^2$, and is thus zero. 

In particular, a class of 
purely gravitational (sourceless) solutions (of $R_{++}=0$) are given by 
\be
H=\sum_{ij} A_{ij}x^i x^j, \;\; tr A=0
\ee
and preserve 1/2 susy $\Gamma_-\epsilon =0$. 

The Aichelburg-Sexl solution is a solution for a point particle (delta 
function source), moving at the speed of light. It is obtained by boosting 
the black hole solution to the speed of light, and taking its mass M to zero,
while keeping $M e^{\beta}=p=$ const. ($\beta$=boost parameter). But a simpler
way to get it is to boost the energy momentum tensor and then solve the 
Einstein equations for the resulting pp wave (thus we have to assume the 
pp wave ansatz, which however turns out to be consistent with the 
energy-momentum tensor). 

A black hole at rest has 
\be
T_{00}= m_0 \delta^{d-2}(x^i) \delta(y)
\ee
and the rest zero. Boosted, one gets 
\be
T_{00}=\frac{m_0}{\sqrt{1-v^2}}\delta^{d-2}(x^i) \delta (y-vt)
\ee
and corresponding $T_{10}$ and $T_{11}$. At the limit, one has 
\be
T_{++}= p \delta ^{d-2}(x^i)\delta (x^+)
\ee

This means that $H(x^+, x^i)=\delta(x^+) \Phi (x^i)$, where (since Einstein's
equation is $R_{++}= 8\pi G T_{++}$)
\be
 \partial _i^2 \Phi (x^i)=-16\pi G p\delta^{d-2}(x^i)
\ee
($\Phi$ is harmonic with source). 

For 4d gravity, $\Phi =-8pG ln \rho$, and 
\be
ds^2= -dudv -4pG ln \rho^2 \delta(u) du^2 +dx^2 +dy^2
\ee
in the notation of \cite{dh} ($\rho^2=x^2+y^2$), 
but the result is easily generalizable 
to any dimension d higher than four: 
\be
\Phi= \frac{16\pi G }{\Omega_{d-3}(d-4)}\frac{p}{\rho^{d-4}}, \;\;\;d>4
\ee
Particles following geodesics in the A-S metric are subject to two effects
\cite{dh}:

It is found that geodesics going along u at fixed v are straight except at 
u=0 where there is a discontinuity
\be
\Delta v =\Phi =-4Gp \;ln \frac{\rho^2}{l_{Pl}^2}
\ee
where the Planck constant $l_{Pl}$ in the ln is conventional
(only relative shifts, $\Delta v_1-\Delta v_2$, have physical meaning). 
That means that one basically 
has two portions of flat space glued together along u=0 with a $\Delta v$ 
shift. The shift can be easily understood by the fact that after a singular 
coordinate transformation, defined later on in (\ref{coordtr}), the metric 
becomes continous. So geodesics are continous in $(u,v)$ coordinates, which 
means they are discontinous in $(\bar{u}, \bar{v})$ coordinates, with the 
above $\Delta v$.  

The second effect is a ``refraction'' 
(or gravitational deflection, rather), where the angles $\alpha$ and $\beta$
made by the incoming 
and outgoing waves with the plane $u=0$ at an impact parameter $\rho=b$ from 
the origin in transverse space satisfy 
\be
cot \alpha +cot\beta =\frac{4Gp}{b}
\ee
(here p is the momentum of the photon creating the A-S wave), and at small 
deflection angles (near normal to the plane of the wave) we have 
\be
\Delta \theta\simeq \frac{4Gp}{b}
\ee
We can understand this also by using the singular coordinate transformation in
(\ref{coordtr}), as 
\be
\Delta(\frac{\partial \bar{\rho}}{\partial\bar{u}})=
 \Delta(\frac{\partial \rho}{\partial u})+\frac{\partial_{\rho}\Phi}{2}
\ee
and $\Delta(\frac{\partial \rho}{\partial u})=0$ (no refraction in $(\rho,u,v)$
coordinates), so 
\be
\Delta(\frac{\partial \bar{\rho}}{\partial\bar{u}})=
\frac{\partial_{\rho}\Phi}{2}
\ee

One can then describe the scattering of two massless particles of very high 
energy \cite{thooft} ($m_{1,2}\ll M_P, Gs\sim 1$, yet $Gs<1$) 
by saying that particle two creates a 
massless shockwave of momentum $p_{\mu}^{(2)}$ and particle one follows a 
massless geodesic in that metric. 
In covariant notation ($v=x^0-x^1\equiv x^-, \tilde{x}^2\equiv \rho^2=x^2
+y^2$),
\be
\Delta x_{\mu}= -2G p_{\mu}^{(2)} log (\tilde{x}^2/C)
\ee
Then particle one comes in with a free wavefunction
\be
\psi_{(-)}^{(1)}= e^{i\tilde{p}^{(1)}\tilde{x}+ip_{-}^{(1)}v+ip_{+}^{(1)}u}
\ee
and becomes (at u=0, just after the shockwave)
\be
\psi_{(+)}^{(1)}=  e^{i\tilde{p}^{(1)}\tilde{x}+ip_{-}^{(1)}(v- 4Gp^{(2)}
log (\tilde{x}^2/C))}
\ee
Then by definition the scattering amplitude is the Fourier transform of this 
wavefunction
\bea
&&{\cal A}(k_-, \tilde{k})=\frac{1}{(2\pi)^3}\int d^2\tilde{x}dv e^
{-i\tilde{k}^{(1)}\tilde{x}-ik_-^{(1)} v}\psi_{(+)}^{(1)}\nonumber\\
&& =\delta(k_-^{(1)}-p_-^{(1)})
\int \frac{d^2\tilde{x}}{(2\pi)^2}e^{i\tilde{x}(\tilde{p}^{(1)}
-\tilde{k}^{(1)})-iGs log \tilde{x}^2}\nonumber\\
&&=-i\delta(k_-^{(1)}-p_-^{(1)})
\int\frac{d^2\vec{b}}{(2\pi)^2}e^{i\vec{q}\vec{b}}e^{i\delta (b,s)}
\eea
where we have expressed ${\cal A}(s,t)$ via an impact parameter transform 
to an eikonal form, with $\delta (b,s)=p_{+}^{(1)}\Delta v=-Gs log b^2$ 
and after doing 
the $d^2\vec{b}=bdbd\theta$ integration one gets 't Hooft's result
\be
{\cal A}=
\frac{1}{4\pi} \delta (k_-^{(1)}
-p_-^{(1)})\frac{\Gamma (1-iGs)}{\Gamma(iGs)}
[\frac{4}{(\tilde{p}-\tilde{k})^2}]^{1-iGs}
\ee
But
\be
4Gp_+^{(1)}p^{(2)}= Gs \;\;{\rm and}\;\; (\tilde{p}-\tilde{k})^2=-t;
\;\;{\cal A}(k_+, \tilde{k})= \delta(k_+^{(1)}-p_+^{(1)})U(s,t)
\ee
and then we get the differential cross-section
\bea
&& U(s,t)=\frac{1}{4\pi}(\frac{4}{-t})^{1-iGs}\frac{\Gamma(1-iGs)}{
\Gamma(iGs)}\nonumber\\
&&\frac{d\sigma}{d^2k}=\frac{4}{s}
\frac{d\sigma}{d\Omega}= 4\pi^2 |U(s,t)|^2= \frac{4}{t^2}(Gs)^2
\eea
which is like Rutherford scattering, as if a single graviton is exchanged.
(with the effective gravitational coupling $Gs$ replacing $\alpha= e^2/4\pi$
of QED) 

The argument is that graviton exchange dominates the amplitude  in this limit, 
for massive particles it takes an infinite time to interact.
Indeed, at large impact parameter there is the natural exponential decay 
of the massive interactions, whereas at small impact parameter the harmonic 
function $\Phi(r)$ diverges, and as the time shift $\Delta v$ is proportional 
to $\Phi$, it diverges as well.  Other 
massless particles can be introduced easily: for example Maxwell interactions
are taken into account just by having a shift (such that at $Gs=0$ we 
recover Rutherford scattering of QED):
\be
Gs\rightarrow Gs +q^{(1)}q^{(2)}/4\pi
\ee

For transplanckian scattering, $Gs\gg 1$, one should take both particles as 
creating shockwaves, and these shockwaves should interact and create a 
black hole. 

The generalization to higher dimensions is now pretty straightforward.
Let's first notice, as Amati and Klimcik did also \cite{ac}, that 
a shockwave metric
\be
ds^2=-dudv +\Phi(x)\delta (u)du^2+d\vec{x}^2
\ee
would shift the geodesics at u=0 by $\Delta v= \Phi $ and the S matrix 
was described by 't Hooft by the Fourier transform of the shifted 
wavefunction, giving essentially 
\be
S=e^{ip_v \Delta v}\equiv e^{ip_-\Phi}
\ee
What we mean is that we can perform an impact parameter transform as in D=4
and get 
\be
i{\cal A}= \int\frac{d^{D-2}\vec{b}}{(2\pi)^{D-2}}e^{i\vec{q}\vec{b}}
(e^{i\delta (b,s)}-1)
\ee
with ($\mu=\sqrt{s}/2=p=$ photon energy and $p_-^{(1)}=\mu$ also)
\be
\delta (b,s)=p_{-}^{(1)}\Phi (b)=\frac{aGs}{b^{D-4}};\;\;\;
\Phi(b)=\frac{16\pi G\mu}{\Omega_{D-3}(D-4)b^{D-4}}
\ee
so $ a= 4\pi/(\Omega_{D-3}(D-4))$. Then one obtains (with $q^2=t$)
\bea
i
{\cal A}&=& \frac{\Omega_{D-4}\Gamma(\frac{D-3}{2})\sqrt{\pi}2^{\frac{D-4}{2}}}
{(2\pi)^{D-2} q^{D-2}}\int _0^{\infty}dz z^{\frac{D-2}{2}}(e^{i\alpha/z^{D-4}}
-1)J_{\frac{D-4}{2}}(z)\nonumber\\
&\equiv &\frac{A}{q^{D-2}}
\int _0^{\infty}dz z^{\frac{D-2}{2}}(e^{i\alpha/z^{D-4}}
-1)J_{\frac{D-4}{2}}(z)
\eea
where the q dependence of the integral comes from $\alpha= a Gs q^{D-4}
=a Gs t^{\frac{D-4}{2}}$ and $z=qb$, and the exponential is $e^{i\delta}$
in general, so for small $\delta$ the bracket in the integral is $i\delta$.
 The integral can also be rewritten as 
\be
\int_0^{\infty}\frac{du}{4-D}u^{-\frac{3D-8}{2(D-4)}}(e^{i\alpha u}-1)
J_{\frac{D-4}{2}}(u^{-\frac{1}{D-4}})
\ee
but we can find no analytic expression for it. At most one can make an 
expansion in $\alpha$ which gives for the integral $=i\alpha c, c=2^{(6-D)/2}
/\Gamma((D-4)/2)$, and so 
\be
{\cal A}\simeq \frac{Gs}{t}\times (\frac{ac \Omega_{D-4}\Gamma(\frac{D-3}{2})
\sqrt{\pi}2^{\frac{D-4}{2}}}{(2\pi)^{D-2}})= \frac{Gs}{\pi t} \frac{1}{
(2\pi)^{D-4}}
\ee
But this is an expansion in $Gs t^{(D-4)/2}$ and so in D=10 we have 
$Gs t^3\ll 1 $, or $g_s (\alpha ' s)(\alpha ' t)^3\ll 1$, certainly satisfied. 
Note also that this result matches in D=4 what one obtains by expanding in 
Gs.

\section{Black hole production via Aichelburg-Sexl wave scattering}

Let us now analyze black hole production in the high energy collision 
of particles ($Gs\gg 1$). We will analyze the collision of two massless 
particles in flat space, in D=4 and $D>4$, first reviewing the treatment 
of Eardley and Giddings \cite{eg}. As noted by 't Hooft and 
analyzed by \cite{eg}, in this regime we have to take into account the 
gravitational field created by both particles, so one has to analyze the 
scattering of two A-S waves. For an estimate of the gravitational 
energy being radiated away in the high energy collision see \cite{cdl}.

As one can imagine, in general, the collision of two gravitational waves 
is a highly nonlinear and nontrivial process, and as such it is hard 
to say anything about the collision region. If we denote by I the region 
$u<0, v<0$ before the collision, by II the region $u>0, v<0$ (after the 
wave at u=0 has passed), III for $u<0, v>0$ (after the wave at v=0 
has passed) and IV for $u>0, v>0$ (the interacting region, after both 
waves have passed), the solution in region IV was calculated in \cite{dp}
only perturbatively in the distance away from the interaction point $u=v=0$.

In the case of sourceless waves (pure gravitational waves), Khan and Penrose 
\cite{kp} and Szekeres \cite{szek} have found complete interacting solutions,
but they don't represent the collision of photons. We will discuss them 
in a next subsection. A general treatment of collision of pure gravitational
waves can be found in \cite{grif}, as well as in \cite{kh,tipler}

\subsection{ Review}

Coming back to the case of the collision of two A-S waves, there is an 
observation, first due to Penrose, and extended by Eardley and Giddings,
which permits one to say that there will be a black hole in the future 
of the collision without actually calculating the gravitational field. 
One can prove the existence of a trapped surface, and then one knows that 
the future of the solution will involve a black hole whose horizon 
will be outside the trapped surface.

An aparent horizon is the outermost marginally trapped surface. The existence 
of a marginally trapped surface thus implies an aparent horizon outside it.
A marginally trapped surface is  defined as a closed spacelike D-2 surface, 
the outer null normals (in both future-directed directions) of which 
have zero convergence. In physical terms, 
what this means is that there is a closed surface whose normal null geodesics 
(light rays) don't diverge, so are trapped by gravity. For a 
Schwarzschild black hole, 
the marginally trapped surface is a sphere around the singularity, that 
happens to coincide with the horizon. 
 
Convergence
 is easier to define in the case of a congruence of timelike geodesics.
For a congruence of timelike geodesics characterized 
by the tangent vector $\xi^a, \xi^a\xi_a=-1$, defining $B_{ab}=\nabla_b \xi_a$
and the projector onto the subspace orthogonal to $\xi^a$, $h_{ab}=g_{ab}
+ \xi_a \xi_b$ (induced metric), the convergence is $\theta=B^{ab}h_{ab}$. 

But we need the case of null geodesics, that is more involved. 
We have to first define the affine parameter $\lambda$ along the curve 
C such that 
\be
\frac{D}{d\lambda} (\frac{\partial}{\partial \lambda})_{C}= \frac{D}{d\lambda
}\xi^a= {\xi^a}_{;b}\xi^b=0
\ee
Then we define a (``pseudoorthonormal'') 
basis for the tangent space, $E_1, E_2, E_3, E_4$, such that $E_4^a=\xi^a$, 
and $
E_3^a=L^a$ is another null vector: $E_3\cdot E_3=E_4\cdot E_4=0, E_1\cdot E_1
=E_2\cdot E_2=1$ and $E_{1,2}$ orthogonal to $E_{3,4}$, but $E_3^a \xi^b g_{ab}
=-1$. And if m,n takes the values 1,2 
in the above basis, then $\theta = \xi_{m;n}g^{mn}$. 
If the geodesics are null, one cannot find an orthonormal basis 
(as in the timelike case), one can only find 
this pseudo-orthonormal basis. Also note 
that by definition, the null geodesics defined by $\xi^a$ are normal to the 
2d surface spanned by $E_1,E_2$, and we are taking the derivative of $\xi$ 
just in those directions.

So to calculate the existence of a marginally trapped surface, we first need 
to find the null geodesics normal to the surface, and then impose that their
convergence is zero.

To calculate the convergence, take the approach from \cite{horava}. The 
convergence is 
\be
\theta = h^{ab}D_{a}\xi_{b};\;\;\; \xi=\xi^{a}\partial_{a}
=\frac{dx^{a}}{d\lambda}\partial_{a}
\ee
for a congruence of null geodesics $\xi_{\mu}$ normal to the surface B, and 
$h_{ab}$ is the induced metric on B. B is spanned by the $E_1, E_2$ of 
before, and contracting with the induced metric is equivalent to contracting
with $g^{mn}$ in the above basis. 

We see that we need to get the form of the geodesics $x^{a}(\lambda)$ 
to proceed. We can impose the fact that the geodesics are null, so 
$(\xi, \xi)=0$, normal to the generators of the surface, $K_i$, 
so $(\xi, K_i)=0$, and also the normalization $(\xi, \partial_t)= -E$ 
(which can be chosen to be $=-1$ for simplicity). 
Note that in \cite{horava} B is a sphere, so the generators are 
$\partial_{\phi_i}$.
Then one calculates 
$x^{a}(\lambda)$ and then $\xi(\lambda(x^{a}))= \xi(x^{a})$, and then 
$h_{mn}=\partial_mX^{a}\partial_nX^{b}g_{ab}$ (where $X^a$ are coordinates
on the surface B),
and finally $\theta= h^{mn} D_m \xi_n$.

Let's apply this procedure to the metric of two colliding general shockwaves
(without specifying for the moment the Aichelburg-Sexl solution for 
$\Phi_i$), one moving in the u direction, and the other in the v direction. 
\be
ds^2=-d\bar{u}d \bar{v} +d\bar{x}^{i2}
+\Phi_1(\bar{x})\delta (\bar{u})d\bar{u}^2
+\Phi_2(\bar{x})\delta (\bar{v})d\bar{v}^2
\ee
After the coordinate transformation 
\bea
&&\bar{u}=u+\Phi_2\theta(v)+v\theta(v)\frac{(\nabla \Phi_2)^2}{4}\nonumber\\
&&\bar{v}=v+\Phi_1\theta(u)+u\theta(u)\frac{(\nabla \Phi_1)^2}{4}\nonumber\\
&& \bar{x}^i=x^i+\frac{u}{2}\partial_i \Phi_1(x)\theta(u)+\frac{v}{2}
\partial_i\Phi_2(x)\theta(v)
\label{coordtr}
\eea
it becomes
\be
ds^2= -dudv +[H_{ik}^{(1)}H_{jk}^{(1)}+H_{ik}^{(2)}H_{jk}^{(2)}-\delta_{ij}]
dx^i dx^j
\ee
where 
\bea
H_{ij}^{(1)}&=& \delta _{ij}+ \frac{1}{2} \partial_i\partial_j \Phi^{(1)} u 
\theta (u)\nonumber\\
H_{ij}^{(2)}&=& \delta _{ij}+ \frac{1}{2} \partial_i\partial_j \Phi^{(2)} v 
\theta (v)
\eea
At zero impact parameter (b=0), and for A-S shockwaves in D=4, we have 
\be
\Phi_1=\Phi_2=-8G\mu ln \bar{\rho};\;\;\; \bar{\rho}=\sqrt{\bar{x}^i\bar{x}^i}
\ee
In general D, but for an A-S wave at b=0, there is a 
D-2 dimensional trapped surface consisting of two disks (balls), 
parametrized by $\bar{x}$, of radius $\rho_c$ in $\bar{\rho}$. 

In complete generality, the surface S is defined as follows.
Take the union of the two null hypersurfaces
$v\leq 0, u=0$ and $u\leq 0, v=0$ with a D-2 dimensional intersection
$u=v=0$, that intersects on its turn S on a D-3 surface C. (a priori, two
D-2 surfaces intersect on a D-4 surface though, more on that later). Then S 
is composed of 
 
``disk'' 1- $\{ v=-\Psi_1(\vec{x}), \;\; u=0\}$, ($\Psi_1=0$ on C)

``disk'' 2- $\{ u=-\Psi_2(\vec{x}),\;\; v=0 \}$  ($\Psi_2=0$ on C).
As we will show, the condition of zero convergence implies that 
\be
\nabla^2 (\Psi_1-\Phi_1)=0
\ee
interior to C. We will see that in the b=0 A-S case, we can actually choose 
$\Psi_1
=\Phi_1, \Psi_2=\Phi_2$ which, with the definition $\theta(0)=1$, means that 
both disks correspond to $\bar{u}=\bar{v}=0$. So we wouldn't see the topology 
in the bar coordinates, we need to go to the unbarred ones to get explicit 
formulas. 

On the first disk, we have 
\bea
ds^2&=&-dudv +d\vec{x}^2 +\frac{1}{2} u\theta (u) (\partial_i\partial_j \Phi)
dx^idx^j+\frac{u^2}{4}\theta(u) \partial_i \partial_k\Phi \partial_j\partial_k
\Phi dx^i dx^j\nonumber\\
&=&-dudv +dx^i dx^j g_{ij}
\eea
and the null geodesics through $\{v=-\Psi(\vec{x}), \;\; u=0 \}$ are 
defined by
\be
\xi =\dot{u}\frac{\partial}{\partial u}+\dot{v}\frac{\partial}{\partial v}
+\dot{x}^i\partial _i
\ee
The tangent generators of the surface are 
\be
K_j^{\mu}=\frac{\partial X^{\mu}}{\partial x^j}
\ee
where $x^i$ are the coordinates on S and $X^{\mu}$ the coordinates on the 
space, but we choose $x^i=X^i$ and so 
\be
K_j^{\mu}=(0, -\partial_j \Psi, \delta_j^i)\rightarrow 
K_j^{\mu}\partial_{\mu}=-\partial_j \Psi \partial_v +\partial_j
\ee
And we have to impose the condition that $\xi$ is null $(\xi, \xi)=0$,
transverse to all the generators: $(\xi, K_i)=0$, and we have to define the 
time direction (in \cite{horava}, that was $(\xi, \partial _t)=-E$, 
where E can be
scaled to 1), in this case $(\xi, \partial_v) =-1$. 

These conditions together fix
\be
\dot{u}=2;\;\;\; \dot{x}^i=-g^{ij}\partial_j \Psi;\;\;\; 
\dot{v}=\frac{1}{2}\partial_i\Psi \partial_j\Psi g^{ij}
\ee
and then we calculate
\be
\xi=-dv-\frac{1}{4}g^{ij}\partial_i\Psi\partial_j\Psi du-\partial_i\Psi dx^i
\ee
and thus 
\be
\theta|_{u=0}=-\nabla^2 (\Psi -\Phi)
\ee
as advertised.

Actually, what we have found is that by imposing $(\xi_1,\partial_v)=-1$, 
we get
\be
\xi_1=-dv-\frac{1}{4}(\nabla \Psi_1)^2 du-\partial_i \Psi_1 dx^i
\ee
but similarly if we impose instead $(\xi_1', \partial_u)=-1$, we get 
\be
\xi_1'=-du -\frac{4}{(\nabla \Psi_1)^2}dv -4\frac{\partial_i \Psi_1}{(\nabla
\Psi_1)^2}dx^i. 
\ee
Then on disk 2, $(\xi_2, \partial_u)=-1$ implies
\be
\xi_2=-du-\frac{1}{4}(\nabla \Psi_2)^2 dv-\partial_i \Psi_2 dx^i
\ee
And these two surfaces intersect on C, thus the normal, $\xi$, has to be 
continous across C. This means that for the A-S wave 
at b=0, when $\Phi_1=\Phi_2$ implying 
$\Psi_1=\Psi_2$, we need to have 
\be
(\nabla \Psi_1)^2=(\nabla \Psi_2)^2=4
\ee

Then in D=4, replacing the explicit form of $\phi$ we get 
\be
\Psi=\Phi =-8G\mu \; ln \;\rho/\rho_c\Rightarrow \rho_c =4G\mu =r_h
\ee
whereas for $D>4$ 
\be
\Psi=\frac{16\pi G\mu}{\Omega_{D-3}(D-4)\rho^{D-4}}\Rightarrow 
\rho_c=(\frac{8\pi G\mu}{\Omega_{D-3}})^{\frac{1}{D-3}}
\ee

In the bar coordinates, both disks correspond as we said to $\bar{u}=
\bar{v}=0$ and $\bar{x}^i=x^i$. But this surface in the bar coordinates is 
just flat (on it, the metric is Minkowski), 
so the area (volume of balls) is just the area of two flat balls 
of radius $\bar{\rho}_c=\rho_c$. The area (volume) of a flat unit D dimensional
ball is $V_{ball,D}
=\Omega_{D-1}/D$, so the total area of the trapped surface 
in D spacetime dimensions (two flat balls) is 
\be
A_{min}(S)=2V_{ball, D-2}\rho_c^{D-2}=\frac{2}{D-2}\Omega_{D-3}\rho_c^{D-2}
\ee
whereas, from the explicit form of the Schwarzschild solution in D dimensions
the horizon radius of a black hole of mass $\sqrt{s}=2 \mu$ is  
\be
r_h=[\frac{32\pi G \mu}{(D-2)\Omega_{D-2}}]^{\frac{1}{D-3}}
\ee
so that the horizon area of the mass= $\sqrt{s}$ black hole is 
\be
A_{Sch}=\Omega_{D-2}r_h^{D-2}\Rightarrow
\frac{A_{min}(S)}{A_{Sch}}=\frac{1}{2}[\frac{(D-2)\Omega_{D-2}}{4\Omega_{D-3}}
]^{\frac{1}{D-3}}\equiv \frac{\epsilon}{2}\label{bharea}
\ee
The area of the trapped surface is smaller than the horizon area of the black 
hole to form (since the horizon is by definition outside the trapped surface),
and we can express the area of the {\em disks (balls)} as the area 
of horizon {\em spheres} that will form,  so $r\leq r_h$, where r is defined 
by Area(S)$=\Omega_{D-2}r^{D-2}$, implying that the mass of the formed black 
hole satisfies
\be
\frac{16\pi G M_{BH}}{(D-2)\Omega_{D-2}}=r_h^{D-3}\geq r^{D-3}=
[\frac{Area(S)}{\Omega_{D-2}}]^{\frac{D-3}{D-2}}\Rightarrow
\frac{M_{BH}}{\sqrt{s}}\geq \frac{1}{2}[\frac{(D-2)\Omega_{D-2}}{2\Omega_{D-3}
}]^{\frac{1}{D-2}}
\ee
(we have put in the explicit form of $Area(S)$ and of $\rho_c$ in terms of 
$\mu =\sqrt{s}/2$).
Both $\frac{A_{min}(S)}{A_{Sch}}$ and  $\frac{M_{BH}}{\sqrt{s}}$ match the 
explicit numbers in \cite{eg}. 

\subsection{ Extension}

In the previous discussion we have already set up the formalism so that 
it is valid for any function $\Phi(\vec{x})$ characterizing the shockwave.
We will be applying this later for different $\Phi$'s. 

Let us now try to extend this for the case of nonzero b in any dimension.
For the b=0, D=4 A-S wave we had 
\be
\Psi=\Phi=-8G\mu \;ln \;\rho/\rho_c
\ee
meaning that $\Psi >0$ for $\rho<\rho_c$. For b=0, $D>4$ we have
\be
\Psi= \Phi-\Phi(\rho=\rho_c);\;\; \Phi=\frac{16\pi G\mu}{\Omega_{D-3}(D-4)
\rho^{D-4}}
\ee
and again $\Psi>0$ for $\rho<\rho_c$. 

For $b>0$, $D\geq 4$ now, we would need both $\psi_1$ and $\psi_2$ to be zero 
on the same surface (curve, for D=4) C, not on two surfaces $C_1$ and 
$C_2$, since then the intersection of $C_1$ and $C_2$ would have D-4 
dimensions (points, for D=4). So we cannot use in D=4 for instance
\be
\Psi_i=\Phi_i=-8G\mu \; ln \; \frac{|\rho-\rho_{0 i}|}{\rho_c}
\ee
That would define two disks in $\bar{x}=x$ with the centers dispaced by b, 
and while each of the disks boundary would be a circle, the two circles 
will intersect in two points, so C would be composed of these two points.

The correct solution, which was explored in \cite{eg} using a self-consistent 
approach (which does not guarantee to find ALL solutions) is that 
$\Psi_i\neq \Phi_i$, just 
\be
\nabla^2 \Psi_i =\nabla^2 \Phi_i\propto \delta (\vec{x}-\vec{x}_{0i})
\ee
which means that $\Psi_1, \Psi_2$ are Green's functions for sources at 
$\vec{x}_{01}, \vec{x}_{02}$
which {\em both} are zero on the same curve C enclosing $\vec{x}_{01}$ and 
$\vec{x}_{02}$. Then one imposes the condition for continuity of the 
null normal $\xi$ which gives
\be
\nabla \Psi_1 \cdot \nabla \Psi_2 =4\label{condition}
\ee
which fixes (together with the previous conditions) the form of C. 

Clearly for very small b (much smaller than $\rho_c$) 
we have that C is well approximated by the boundary 
(envelope) of the union of the two disks. We will assume that in D=4 
the two points $C_1$ and $C_2$ 
(intersection of the two circles=boundaries of disks) are still part of the 
curve C even at b large, which seems like a reasonable assumption, though not 
well justified. Let's see what we can deduce out of it. Clearly the two 
sources will be inside C, so we will therefore assume that the curve C is 
outside the paralelogram made from $C_1, C_2, x_{01}, x_{02}$. We still 
call the distance between $C_{1,2}$ and 
$x_{01}, x_{02}$ (radius of the circles) $\rho_c$, and if we then 
impose (\ref{condition}) on $\Psi_1=\Phi_1$ and $\Psi_2=\Phi_2$ (which are 
still good Green's functions for the circles that both pass through the
two points $C_1, C_2$, even if they are
not for the whole curve C) we get the equation ($cos \;
\theta /2= \sqrt{\rho_c^2
-b^2/4}/\rho_c$)
\be
|\nabla \frac{\Psi_1}{2}| \cdot |\nabla \frac{\Psi_2}{2}| 
cos \theta=1 \Rightarrow \frac{R_s^2}{\rho_c^2}(1-\frac{b^2}{2\rho_c^2})=1
\ee
(where $R_s=4\mu G$) which gives the value of $\rho_c$ as  
\be
\rho_c^2=\frac{R_s^2}{2}(1+\sqrt{1-\frac{2b^2}{R_s^2}})
\ee
We can check that if b=0 we reproduce the known result of $\rho_c=R_s$.
This formula
means that the maximum impact parameter for which we can have a black 
hole forming within this approximate formalism is $b_{max}= R_s/\sqrt{2}
=4G\mu/\sqrt{2}$ and 
the minimum radius is $\rho_{c, min}= \rho_c(b_{max})=R_s/\sqrt{2}=b_{max}$, 
and the area of the trapped surface satisfies
\be
S \geq \sqrt{b^2 \rho_c^2-\frac{b^4}{4}}= \frac{b}{\sqrt{2}}\sqrt{R_s^2(1+
\sqrt{1-\frac{2b^2}{R_s^2}})-\frac{b^2}{2}}\equiv S_{min}
\ee
so that $S_{min}$ at the maximum b is $S_{min}= R_s^2\sqrt{3}/4= 4\sqrt{3}
(\mu G)^2$. 

Comparing now with the results of \cite{eg} we have that $b_{max}= 4G\mu
/\sqrt{2}\simeq 2.83 G\mu$  is 
smaller than their result of $3.219 G \mu$. Since $b_{max}<R_s$ it is even 
physically acceptable (we would have a problem if it would be bigger). 
As for the estimate of the area of the trapped surface, $S_{min}=4\sqrt{3}
 (\mu G)^2$,
it is sensibly smaller than the result of \cite{eg} which can be found to be
(replacing the value of their parameter $a_{max}$ in the formula for the area)
$40.852 (\mu G)^2$, so we have a much more conservative estimate.

But the advantage is that this procedure can be now easily extended in higher 
dimensions. 

Indeed, in $D>4$ ,
\be
\vec{\nabla} \Phi=-\frac{16\pi \mu G}{\Omega_{D-3}}\frac{\vec{x}}{\rho^{D-2}}
\ee
and so the condition $\vec{\nabla}\Phi_1\cdot \vec{\nabla}\Phi_2=4$ implies
\be
(\frac{\epsilon R_s}{\rho_c})^{2D-6}(1-\frac{b^2}{2\rho_c^2})=1
\ee
where $R_s\equiv r_h$ is the horizon radius of the black hole, $\epsilon $ 
is defined in (\ref{bharea}) and this equation can be rewritten as 
\be
f(x)=4x^{D-2}-4\alpha x +2 b^2 \alpha =0 \label{xeq}
\ee
where $x=\rho_c^2$ and $\alpha =(\epsilon R_s)^{2D-6}$.  We can easily
find the maximum value of the impact parameter b from it. Since 
$\rho_c=\sqrt{x}$ is the biggest of the solutions to the equation (\ref{xeq}),
we impose that $f(x_0)\leq 0$, where $x_0$ is the highest root of $f'(x_0)=0$.
This condition implies
\be
b^2\leq 2[\frac{\alpha}{D-2}]^{\frac{1}{D-3}}\frac{D-3}{D-2}= 
\frac{2(\epsilon R_s)^2}{ [D-2]^{\frac{D-2}{D-3}}}(D-3)
\ee
We can check that indeed in D=4 we recover the result $b_{max}=R_s/\sqrt{2}$, 
since then $\epsilon=1$.
In D=5, that means $b\leq 0.9523 R_s< R_s$. 
We can also calculate the lower limit on the 
area of the trapped surface as before, except that now the area of the 
paralelogram $C_1, C_2, x_{01}, x_{02}$ is replaced by the volume of a 
``surface of revolution'' in D-4 transverse directions around the axis $x_{01},
x_{02}$. The geometry in higher dimensions is more complicated, 
but for D=5, this 
is just two cones glued on their bases, of height h=b/2 
and base radius $\rho_c 
\cos \theta/2$ , and so ``$S_{min}$'' (volume of the cones) is 
\be
2\frac{S_0h}{3}= \frac{\pi b}{3}(\rho_c^2 -\frac{b^2}{4})
\ee

In conclusion, we have set up a formalism for shockwaves in which we can 
calculate trapped surfaces at b=0 and to some degree at nonzero b, for a
general shockwave form. 

\subsection{Collision of sourceless waves}

We have seen that for A-S-type waves colliding, in general we get 
a trapped surface in the future of the collision, which indicates a 
black hole horizon being formed. From this, we conclude that a black 
hole is formed in the high energy collision of two high energy photons 
(massless particles, with an energy momentum source). 

But what happens if two sourceless waves (gravitational solutions to the pure 
Einstein's equations) collide? We would expect to be able to associate this 
phenomenon with the collision of two gravitons, in which case we would expect 
to create a black hole in the collision. It is in fact true that there is a
theorem stating that a singularity will form in the future of a collision 
of two sourceless waves \cite{kh,tipler}. 
It is also a theorem that for Einstein gravity 
in flat space, a singularity cannot be naked, so we would expect to be 
able to find a trapped surface, indicating the formation of a horizon in a 
sourceless wave collision. 

Unfortunately, we will see that this is not so, and we will speculate on 
why, after we see the problem. 

Khan and Penrose found a solution \cite{kp} describing the 
(head-on, zero impact parameter) collision of two source-free gravitational 
pp waves of the type
\be
ds^2=-dU(2dV+(X^2-Y^2)h(U)dU)+dX^2+dY^2
\ee
with $h(U)=\delta(U)$ (the function $\Phi$ 
defined before satisfies the source-free 
equation $\partial_i^2 \Phi=0$ solved by $\Phi=-X^2+Y^2$). 
After the coordinate transformation 
\be
U=u,\;\;\;V=v+x^2/2 FF'+y^2/2 GG';\;\;\; X=xF;\;\;\;Y=yG
\ee
with $F''=-Fh, G''=Gh$, solved by
$F=1-u\theta(u), \;G=1+u\theta(u)$ ($F'=\theta (u)$, as $u\delta (u)=0$, and 
thus also $F\delta(u)=\delta (u)$), we get the wave in the form
\be
ds^2=-2dudv+F^2 dx^2+G^2 dy^2
\label{fgwave}
\ee
The collision will involve two such waves, one in u and the other in v, at 
zero impact parameter (b).
Thus the colliding wave solution of Khan and Penrose is
\be
ds^2=-\frac{2t^3dudv}{rw(pq+rw)^2}+t^2(\frac{r+q}{r-q})(\frac{w+p}{w-p})dx^2
+t^2(\frac{r-q}{r+q})(\frac{w-p}{w+p})dy^2
\ee
where 
\be
p=u\theta(u);\;\;q=v\theta(v);\;\;r=\sqrt{1-p^2};\;\;w=\sqrt{1-q^2};\;\;
t=\sqrt{1-p^2-q^2}
\ee
In the region $u\ge 0, v<0$ (before the coming of the second wave), we can 
check that the Khan-Penrose solution becomes 
\be
ds^2= -2dudv+ (1+p)^2dx^2+(1-p)^2 dy^2
\ee
that is, of the sourceless wave form (\ref{fgwave}), and we see that there 
is a coordinate singularity at u=1. 
Then in the collision region $u>0,v>0$ we have 
\bea
&&ds^2=-2\sqrt{\frac{1-u^2-v^2}{(1-u^2)(1-v^2)}}
\frac{(\sqrt{(1-u^2)(1-v^2)}-uv)}{(\sqrt{(1-u^2)(1-v^2)}+uv)}dudv
\nonumber\\&&
+\frac{(\sqrt{1-u^2}+v)^2(\sqrt{1-v^2}+u)^2dx^2+(\sqrt{1-u^2}-v)^2
(\sqrt{1-v^2}-u)^2dy^2}{1-u^2-v^2}
\eea
Putting v=0 we get back to the sourceless wave solution (\ref{fgwave}).
And \cite{kp} found that in the collision region, the line $u^2+v^2=1$ has a 
scalar curvature singularity. We can calculate that for $u^2+v^2=1-\epsilon$
the metric is 
\be
ds^2=\epsilon[\frac{\sqrt{\epsilon}du^2}{2u^2v^4}+(\frac{4uv}{\epsilon})^2dx^2
+(\frac{\epsilon}{4uv})^2dy^2]
\ee
so clearly the metric is singular, but there doesn't seem to be any good way
to define a finite area of the singularity. Indeed, at u=fixed, v=fixed, 
\be
dS=ds_x\cdot ds_y=\epsilon dxdy
\ee

So we can't calculate this way a minimum on a horizon area of a black hole 
that would probably form. 

But  we can still try to apply the formalism of Eardley and Giddings
and calculate the area of a trapped surface that we assume will (should)
form. Indeed, the 
individual gravitational waves that collide are still of the general form 
used in the previous subsection. The only difference is that 
instead of $\Phi=-8G\mu \; ln \;\rho/\rho_c$ we have 
\be
\Phi= -\frac{X^2-Y^2}{\rho_c};\;\;\;(\rho_c=4G\mu)
\ee
where we have rescaled U and V to introduce the dimensionful parameter 
$\rho_c$ describing the strength of the wave. The problem is though that 
in order to be able to choose $\Psi = \Phi$ like we did for the b=0 A-S
wave collision, 
$\Phi$ would have to be zero on the curve C at its boundary, so that  $\rho=
\rho_c$ as we shall see. But $\Phi=0$ for $X=\pm Y$ (we could shift $\Phi$ 
by a constant, and then C would be a hyperbola), so that we can't actually 
choose $\Psi= \Phi$. 

Thus we could only use the above function for $X=\pm Y$, 
and these are four points ($X=\pm \rho_c/\sqrt{2}, X=\pm Y$) 
that would presumably lie on the curve C if there would be such a curve. 

However, the correct treatment would involve solving the 2d Green's 
function (``electric potential'') for 
the Laplace equation $\nabla^2 \Psi=0$
with Dirichlet boundary conditions $\Psi=0$ on a curve C 
where $\nabla \Psi$ (the ``electric field'')  has unit norm, and this 
is impossible!

Thus we seem to have proven that the assumption of a trapped surface is 
in fact wrong!

So there  really seems to be no way of obtaining a trapped surface in 
the Khan-Penrose solution, even though we do obtain a singularity!

This is a most bizzare situation in itself, which could be perhaps saved 
by the fact that in such singular spacetimes the usual censorship
theorems don't apply,
but correlated with the expectation that the Khan-Penrose solution should  
describe graviton-graviton scattering, this is really puzzling.

One could perhaps think that the Khan-Penrose metric is not the correct 
sourceless wave to describe graviton-graviton scattering. After all, 
there is a plethora of sourceless wave scattering solutions, as reviewed 
to certain extent in \cite{grif}. 

One can analyze their behaviour though (we will not do it here explicitly)
and convince oneself that these solutions do not describe graviton scattering.
The simplest of them is the Szekeres solution in \cite{szek}, which has the 
same singularity structure as the Khan-Penrose solution, but is described 
by a function $\Phi$ of the type $\Phi(\tilde{u})= f(\tilde{u})
\theta(\tilde{u})$
as opposed to the delta function profile $\Phi(\tilde{u})=\delta(\tilde{u})$
of the incoming waves in the Khan-Penrose solution. The rest of the possible 
solutions are even more complicated, and they really describe the collision 
of realistic gravitational waves, as opposed to the idealized delta function
waves of the Khan-Penrose solution. Therefore the Khan-Penrose 
solution is the only one that 
can claim to represent the collision of two (idealized) gravitons. 

Of course, all these solutions were in 4d general relativity. 

Gutperle and Pioline \cite{gp} set out to generalize these solutions to 
2n+2 dimensions and to add p-form field strength to it, the ultimate goal 
being to scatter 2 maximally susy pp waves of IIB, or rather a 
shockwave generalization of it. They fall kind of 
short of the goal. The first try at the generalization gives exact solutions 
which however do not satisfy appropriate boundary conditions: the incoming 
waves are different from the Khan-Penrose and Szekeres profiles. 

A perturbative attempt near the lightcone (or for the strength of a 
wave much smaller than the other) produces a higher dimensional solution, 
as well as the p-form generalization. 

Then, \cite{ccfl} also produce some generalizations of this type
(see also \cite{chen,cz}), with 
better singularity structure, but they don't analyze the incoming 
waves in Brinkman form, so it is not clear what they correspond to.

In conclusion, the Khan-Penrose solution is the only one that has a 
chance of describing the collision of two idealized gravitons, and we 
seem to obtain the existence of a naked singularity (no black hole!) in 
the future of the collision. A good explanation of this paradox is still 
lacking.

\section{String corrections}

We will now try to apply the previously derived formalism to shockwave
metrics that incorporate string corrections to the high-energy scattering 
of two photons. 

There are two such formalisms. One is due to Amati and Klimcik \cite{ac}, and
the other due to Amati, Ciafaloni and Veneziano \cite{acv93} (see also 
\cite{amati,acv87,acv89}).

The approach by \cite{acv93} involves writing down an effective 
shockwave metric from which one can calculate an S matrix, which then 
is matched with a string-corrected S matrix (string calculation). 

The S matrix is defined as $exp (iS/\hbar)$ where the action is a function 
of the  classical effective metric, with a source coupling to an external 
$T_{\mu\nu}$. Namely, the S matrix is 

\be
S_{eff}(b, E)= < e^{iA(h_{\mu\nu})/\hbar}>_{tree}= e^{iA(h^{\mu\nu}_{cl})/\hbar
}
\ee
where 
\be
A(h^{\mu\nu})=\int d^4x [L_{eff}(h^{\mu\nu})+T_{\mu\nu}h^{\mu\nu}]
\ee
is the action evaluated on its classical solution with sources given by 
two shockwaves at x=0 and x=b
\be
T_{--}=kE\delta(x^-)\delta^2(\vec{x});\;\;\;\; T_{++}=kE\delta(x^+)\delta^2 
(\vec{x}-\vec{b})
\ee
and $L_{eff}$ is an effective Lagrangian by Lipatov
\cite{lipatov}. Amati et al. \cite{acv93}
showed as we mentioned that this 
calculation reproduces the result for the string correction to the scattering
matrix S.

The string-corrected A-S type metric 
obtained in \cite{acv93}
for D=4 can be expressed in terms of a $\Phi$ of the form 
\be
\Phi= \Phi^{(0)}(A-S)+\Phi^{(1)}=
kE[-\frac{1}{2\pi} log \frac{|z|^2}{L^2}+R_s^2 a^{(1)}(z)]
\ee
where
\be
a^{(1)}(z)=\frac{1}{4\pi}[\frac{1}{|z|^2}|1-\frac{z}{b}|^2 log 
|1-\frac{z}{b}|^2+\frac{1}{bz}+\frac{1}{bz^*}-\frac{1}{b^2}log \frac{L^2}{b^2}]
\ee
and $kE=8\pi G \mu$ ($k=8\pi G, E\equiv \mu$), so the coefficient of the 
log term  in $\Phi^{(0)}$ is $-R_s$. Here 
 $z=x^1+ix^2$ are complex transverse coordinates  so that $|b-z|^2
=(b-x_1)^2+x_2^2$, etc. 

Then 
\bea
&&\partial_1\Phi^{(1)}= \frac{R_s^3}{b(x_1^2+x_2^2)^2}[(x_1^2-x_2^2-bx_1)log 
\frac{(b-x_1)^2+x_2^2}{b^2} +\frac{x_1}{b}(x_1^2+x_2^2-2b x_1)]\nonumber\\
&&\partial_2\Phi^{(1)}=\frac{R_s^3 x_2}{b(x_1^2+x_2^2)^2}[(2x_1-b)log 
\frac{(b-x_1)^2+x_2^2}{b^2}+\frac{1}{b}(x_1^2+x_2^2-2bx_1)]
\eea
But the problem is that the string-corrected metric is only valid for $b> R_s$ 
(when no black hole forms yet), whereas we want to have a perturbation in b
small.

We have tried to just plug in this metric 
in the continuity condition 
$\vec{\nabla}\Psi_1 \cdot \vec{\nabla} \Psi_2=4$, and treat it perturbatively 
in $R_s/b$ as in \cite{acv93}, but one gets corrections of the order 
$R_s^2/\rho_c^2$, and then in a perturbative solution, when $\rho_c$ is 
replaced by the first order solution which is of $o(R_s)$, the corrections 
are of order one. Thus this perturbation is useless. 

But one can still do a small calculation, namely to take the corrected metric 
(with b nonzero and moreover $>R_s$) and see what it does to the continuity 
condition for the collision at b=0 of two 
non-corrected metrics, namely $(\nabla \Phi)^2=4$, 
now with 
\be
\partial_i \Phi=-2R_s\frac{x_i}{\rho^2}+\partial_i \Phi^{(1)}
\ee
and expand in $R_s^2/b^2$. This will not be very relevant (since the 
corrections dissappear in the $b\rightarrow \infty $ limit, which is not 
what we want), but just to see
what kind of effect string corrections have. We obtain
\bea
\rho_c^2&=& R_s^2[1-\frac{R_s^2}{b\rho_c^2}A+\frac{R_s^4}{b^2\rho_c^4}[
(\rho_c^2-x_1^2)log^2(\frac{\rho_c^2-2bx_1+b^2}{b^2})+A]\nonumber\\
A&=&(\frac{\rho_c^2-2bx_1}{b}+
(x_1-b)log \frac{\rho_c^2-2bx_1+b^2}{b^2})
\eea
where we should have $x_1^2+x_2^2=\rho_c^2$, but obviously since we have used 
an asymmetric solution (where b is a distance on the $x_1$ axis), the solution
we get for $\rho_c(R_s)$ also depends of our choice for $x_1, x_2$, so
$\rho_c=\rho_c(x_1, x_2)$.

The above solution is exact, but we only need to expand in $\rho_c/b$. 
Expanding to the first two nontrivial orders we get (after some algebra)
\be
\rho_c^2
\simeq R_s^2[1+\frac{x_1}{b^3}(\frac{x_1^2}{3}+x_2^2)+\frac{1}{2b^4}
(8x_1^2x_2^2-R_s^4+4x_1^2R_s^2-\frac{8}{3}x_1^4)...]
\ee
The first correction is proportional to $x_1$ (times a positive quantity), 
so when we calculate the area of the curve $\rho_c(x_1,x_2)$, the positive 
contribution for $x_1>0$ will cancel against the negative one for $x_1<0$.
So we need to turn to the next correction to see whether or not the area 
increases. 

Defining 
\be
f(x_1^2)=8x_1^2x_2^2-R_s^4+4x_1^2R_s^2-\frac{8}{3}x_1^4
\simeq 12 x_1^2 R_s^2-\frac{32}{3}x_1^4-R_s^4
\ee
for $y=x_1^2/R_s^2$ between 0 and 1, we can check that the function is positive
for $y>0.09$ (most of the domain), 
so a simple estimate shows that the area of the trapped surface
will indeed increase.

But after so many approximations it is not clear this still is relevant.

We turn instead to the approach of Amati and Klimcik \cite{ac}.

Amati and Klimcik \cite{ac} first generalize the 't Hooft and 
Dray and 't Hooft calculation, as we explained in section 2. A shockwave metric
\be
ds^2=-dudv +\Phi(x)\delta (u)du^2+dx^2
\ee
would shift the geodesics at u=0 by $\Delta v= \Phi$ and the S matrix 
was described by 't Hooft by the Fourier transform of the shifted 
wavefunction, giving essentially 
\be
S=e^{ip_v\Delta v}
\ee
 
In string theory, the 't Hooft scattering in the shockwave background gives 
(for an open string$\rightarrow$ photon)
\be
\Delta v=\frac{1}{\pi}\int_0^{\pi}\Phi(X(\sigma, 0))d\sigma
\ee
and the S matrix is defined as acting on creation/annihilation operators 
as $S^+a_{in}S=a_{out}$. Then 
\be
S=e^{\frac{ip_v}{\pi}\int_0^{\pi}d\sigma \Phi(X^u(\sigma, 0))}
\ee
This matches the resummed string calculation of \cite{acv87} if 
\be
\Phi(y)=-q^v\int_0^{\pi}\frac{4}{s}:a_{tree}(s, y-X^d(\sigma_d,0)):
\frac{d\sigma_d}{\pi}
\ee
where 
$\frac{2p_vq^v}{s}=-1, b=x^u-x^d$ and $\hat{X}^u, \hat{X}^d$ are nonzero-modes.
Here the indices u,d refer to ``up'' and ``down'', necessary when we evaluate 
$\Phi(X(\sigma, 0))$.

We note that here b refers just to a parameter in the calculation of the 
shape of one modified A-S metric. We haven't reached the scattering of 
two A-S type waves yet. In that case, we will denote the impact parameter 
of the two waves by B, to avoid confusion. 

Then we match with the S matrix obtained by resumming string diagrams, 
\be
S=exp [2i \int_0^{\pi}:a_{tree}(s, b+\hat{X}^u(\sigma_u,0)-\hat{X}
(\sigma_d,0)):\frac{d\sigma_ud\sigma_d}{\pi^2}]
\ee
and the tree amplitude is 
\be
a_{tree}(s,b)= \frac{G_Ns}{2\pi^{D/2-2}}b^{4-D}\int_0^{b^2/(Y-i\pi/2)/4}
dt e^{-t} t^{D/2-3}
\ee
($Y=\alpha 'log s$). Then  $\Phi(y)$ 
becomes the function for
the A-S wave at $y\gg \sqrt{Y}= \sqrt{\alpha ' log s}
$. So S is dominated by graviton exchange at large b (Aichelburg-Sexl) and 
by absorbtion at small b. 

As a first approximation, we can neglect all string oscillators in $\Phi(y)$
and obtain 
\be
\Phi(x)= -\frac{4q^v}{s}
a_{tree}(s,x)
\ee
where $a_{tree}(s,x)$ is $a_{tree}(s,b)$ (impact parameter space), and 
becomes equal to the A-S result at large b. We can rewrite it also as
(g= gauge coupling) 
\be
a_{tree}(b,s)=\frac{\alpha 'g^2s}{16\pi}\frac{1}{(4\pi \bar{Y})^{D/2-2}}\int 
_0^1 d\rho \rho^{D/2-3}e^{-\frac{b^2}{4\bar{Y}}\rho}
=\frac{g^2s\alpha '}
{16\pi}\frac{1}{\pi^{D/2-2}b^{D-4}}\int_0^{\frac{b^2}{4\bar{Y}}}
dt e^{-t}t^{D/2-3}
\ee
where $\bar{Y}= \alpha ' log (-is)=Y-i\pi\alpha ' /2$. 
We are only interested in the real 
part of $a_{tree}$, as it is the only one that we can use in the classical 
gravitational wave scattering calculation. It is obtained jut by replacing
$\bar{Y}$ with Y. For $b^2\gg Y$, we obtain 
\bea
Re \;a_{tree}(s,b)&\simeq&
\frac{g^2s}{16\pi}\frac{\alpha '}{\pi^{D/2-2}b^{D-4}}(\Gamma(D/2-2)-
e^{-\frac{b^2}{4Y}}(\frac{b^2}{4Y})^{D/2-3}(1+(D/2-3)\frac{4Y}{b^2})+...)
\nonumber\\
&=& \frac{g^2s\alpha '}{16\pi}(\frac{2}{\Omega_{D-5}b^{D-4}}+...)
\eea
whereas for $b^2\ll Y$ we get 
\be
Re \; a_{tree}(s,b)\simeq 
\frac{g^2s\alpha '}{16\pi}\frac{2}{(4Y)^{D/2-2}}
(\frac{1}{D-4}-\frac{b^2}{4Y(D-2)}+...)
\ee
One need just repeat the Eardley-Giddings-type calculation now, as we 
have set it up in the previous section. 

The regime we are working in is small g, large $G_Ns=g^2\alpha ' s/(8\pi)$. 
Since $R_s^2= 4 G_N^2 s= g^4{\alpha '} ^2 s/(4\pi)^2$ and $Y=\alpha ' 
log (\alpha ' s)$, 
\be
\frac{R_s^2}{Y}=\frac{g^2}{log (\alpha ' s)}\frac{g^2\alpha' s}{(4\pi)^2}
\ee
can still be arbitrary, in particular it can be very large.  Since 
to first order $b_{max}=\rho_c=R_s$ (for Aichelburg-Sexl), and at large b 
the metric is A-S plus corrections, in the regime $R_s^2/Y\gg 1$ we can 
use the large b  ($b^2/Y\gg1$) expansion of $\Phi(b)$.

Then in D=4 
we get, with $p^uq^v= s\Rightarrow q^v=\sqrt{s}$ (with the choice 
$p^u=q^v$ due to center of mass scattering, with equal strength shockwaves 
scattering)
\be
\Phi(b)= -\frac{g^2\sqrt{s}}{4\pi}\alpha '(2\;log \; \frac{b}{R_s}- 
e^{-\frac{b^2}{4Y}}
(\frac{b^2}{4Y})^{-1}+...)= -R_s(2\;log \; \frac{b}{R_s}- e^{-\frac{b^2}{4Y}}
(\frac{b^2}{4Y})^{-1}+...)
\ee
Then the condition for the trapped surface appearing in the 
scattering of two Amati-Klimcik waves at zero impact parameter, $(\nabla 
\Phi)^2=4$ gives 
\be
b_{max}=\rho_c\simeq R_s(1+(1+\frac{4Y}{R_s^2})e^{-\frac
{R_s^2}{4Y}})
\ee
(for $b^2/Y\gg1$, so $R_s^2/Y\gg 1$) thus increases, 
so the area of the formed black hole  also increases (since the 
black hole area is proportional to $\rho_c^2$) .
The area of the trapped surface giving the bound on the horizon area is
$S_{min}=2\pi \rho_c^2= 4\pi r_h^2$ and $r_h=2M_{bh}G$, so 
\be
M_{bh}=\frac{\rho_c}{2\sqrt{2}G}
\ee
also increases.

At nonzero impact parameter of the two Amati-Klimcik waves, parameter 
denoted by B as we mentioned (to avoid confusion with the b that was 
used previously),
applying the same approximation for finding $\rho_c$ as was used
in the flat space A-S case, the normal continuity 
 condition is $\partial_i \Phi_1\cdot \partial_i\Phi_2
=4$, so $\partial_i \Phi(\vec{x}-\vec{x}_1)\cdot \partial_i \Phi(\vec{x}-
\vec{x}_2)=4$, so we only get an extra factor of 
\be
\cos ^2 \theta=1-\frac{B^2}{2\rho_c^2}
\ee
to the condition, which thus gets modified to 
\be
\frac{\rho_c}{R_s}=\sqrt{1-\frac{B^2}{2\rho_c^2}}(1+
e^{-\frac{\rho_c^2}{4Y}}+...)
\ee
solved perturbatively by
\be
\frac{\rho_c^2
(R_s, B)}{R_s^2}=\frac{1}{2}
(1+\sqrt{1-\frac{2B^2}{R_s^2}+8(y_0-\frac{B^2}{2R_s^2})
e^{-\frac{R_s^2}{4Y}y_0}})+...;\;\;y_0=\frac{1+\sqrt{1-\frac{2B^2}{R_s^2}}}{2}
\ee
which means that  $B_{max}=R_s/\sqrt{2}(1+e^{-R_s^2/(8Y)})$.

Finally, let us see what happens if $R_s^2/Y\ll 1$. At first, we would guess 
that we can use the small b expansion of the metric $b^2/Y\ll 1$, for which 
\be
\Phi(b)=-2R_s(\frac{1}{D-4}-\frac{b^2}{4Y(D-2)}+...)
\ee
But if we plug it into the continuity equation for getting $\rho_c$, $(\nabla
\Phi)^2=4$, we would get $\rho_c=4Y/R_s$ to first order, meaning that 
$\rho_c^2/(4Y)= 4Y/R_s^2$, 
that is we would seem to be in the opposite regime, so the 
perturbation expansion used was invalid! The solution is of course that 
$R_s^2/Y \ll 1$ will correspond to $\rho_c^2/Y\sim 1$, so we would need to 
use the  full solution, which however is difficult to handle. 

But in any case we can say that for $R_s^2/Y\ll 1$, classically (A-S wave) 
we have $\rho_c=R_s$, but in string theory we get $\rho_c\sim \sqrt{Y}\gg 
R_s$, so we have a great increase in the area of the black hole formed, thus 
it is natural to assume the cross section will also increase.

\section{Randall-Sundrum-type models}

The next application of the black hole creation formalism is to see 
what kind of corrections appear if we have the black hole being created
in a physical setting, namely for a Randall-Sundrum scenario for 
low Planck scale. Emparan \cite{emp} found an A-S-type wave in the 
background of the one brane RS scenario, and analyzed the scattering 
a la 't Hooft in this wave. 

Here we will try to see how the addition of the RS background 
affects the Eardley-Giddings calculation for the flat space black 
hole creation. We will keep the wave on the brane, as in the Emparan 
calculation. 

\subsection{A first attempt- applying the formalism}

The solution for an A-S-type wave in the RS background is 
\be
ds^2=dy^2 +e^{-2|y|/l}(-dudv+dx^idx^i+h_{uu}(u, x^i, y)du^2)
\ee
where 
\be
h_{uu}= \frac{4G_{d+1}}{(2\pi)^{(d-4)/2}}p\delta(u) \frac{e^{d|y|/(2l)}}
{r^{(d-4)
/2}}\int_0^{\infty}dq q^{(d-4)/2}J_{(d-4)/2}(qr)\frac{K_{d/2}(e^{|y|/l}lq)}{
K_{d/2-1}(lq)}
\ee
which is a solution of Einstein's equation with 
$t_{uu}=2\pi p \delta(q_0+q_1)$. Yet another form for the metric is
\bea
&&e^{-2|y|/l}h_{uu}(u,r,y)= -4G_4 p\delta(u)[e^{-2|y|/l}log\frac{r^2}{l^2}
\nonumber\\&&-\frac{2l}{
\pi} \int_0^\infty dm K_0(mr) \frac{Y_1(ml)J_2(mle^{|y|/l})-J_1(ml)Y_2(ml e^{
|y|/l})}{J_1^2(ml)+Y_1^2(ml)}]
\eea
which means that on the brane (y=0)
\be
h_{uu}(u,r,y=0)=-4G_4p\delta(u)[\log\frac{r^2}{l^2}-\frac{4}{\pi^2}
\int_0^\infty
\frac{dm}{m}\frac{K_0(mr)}{J_1^2(ml)+Y_1^2(ml)}]
\ee
The Einstein tensor for this solution is linear in $h_{uu}$, and thus
even though 
this is found as a solution to the linearized equations of motion, it is also 
an exact solution.

At large distances, $r\gg l$, 
\be
h_{uu}(u,r;y=0)
=-4G_4p\delta(u)[log \frac{r^2}{l^2}-\frac{l^2}{r^2}+\frac{2l^4}{r^4}
(log \frac{r^2}{l^2}-1)+...]
\ee
whereas at small distances $r\ll l$,
\be
h_{uu}(u,r;y=0)
=-4G_4p\delta(u)[-\frac{l}{r}+\frac{3}{2}log \frac{r}{l}+\frac{3r}{8l}
+...]
\ee

We can use  the formalism developped previously, since the solution can 
also be expressed as just a modification of the $\Phi$ function.
Now we can at least calculate the zero impact parameter (b) 
values of $S_{min}$ (the area of the trapped surface) and the mass of the 
corresponding black hole. We can also estimate the nonzero b parameter
values of $\rho_c(R_s), b_{max}, S_{min}$. 

The new function $\Phi$ is now
\be
\Phi (u,\rho,y=0)=-R_s[\log\frac{\rho^2}{l^2}-\frac{4}{\pi^2}\int_0^\infty
\frac{dm}{m}\frac{K_0(m\rho)}{J_1^2(ml)+Y_1^2(ml)}]
\ee
which means that 
\be
\partial_i \Phi= -R_s \frac{x_i}{\rho}[\frac{2}{\rho}-\frac{4}{\pi^2}
\int_0^{\infty}dm \frac{K_0'(m\rho)}{J_1^2(ml)+Y_1^2(ml)}]
\ee
and thus imposing the continuity of the normal condition $(\partial_i \Phi)^2
=4$ and rescaling the variables by $R_s$ we get the integral equation for 
$\rho_c$
\be
\frac{\rho_c}{R_s}=1-\frac{2\rho_c/R_s}{\pi^2}\int_0^{\infty}dy
\frac{K_0'(y\rho_c/R_s)}{J_1^2(yl/R_s)+Y_1^2(y l/R_s)}
\label{rhoc}
\ee

As before, the area of the trapped surface is the area of two disks, so 
it is 
\be
S_{min}=2\pi \rho_c^2=4\pi r_h^2
\ee
where $r_h$ is the horizon radius of the formed black hole, and $r_h=2GM_{bh}$,
so
\be
M_{bh}=\frac{\rho_c}{2\sqrt{2}G}
\ee

We can use the expansion for $\rho\gg l$ and $\rho\ll l$ to calculate the form 
of $\rho_c$ from the equation (\ref{rhoc}), for $R_s\gg l $ and $R_s\ll l$.
For $R_s \gg l$ we have  
\be
\Phi=\Phi^{(0)}+\Phi^{(1)};\;\;\; \Phi^{(1)}\simeq R_s[\frac{l^2}{\rho^2}
-\frac{2l^4}{\rho^4}(ln \frac{\rho^2}{l^2}-1)+...]
\ee
and thus imposing $(\partial_i \Phi)^2=4$ we get 
\be
\rho_c^2\simeq R_s^2[1+\frac{2l^2}{R_s^2}-\frac{l^4}{R_s^4}(8\;ln 
\frac{R_s^2
}{l^2}-13)]
\ee
For $R_s \ll l$ we get 
\be
\Phi=-R_s[-\frac{l}{\rho}+\frac{3}{2}ln \frac{\rho}{l}+\frac{3\rho}{8l}+...]
\ee
and then 
\be
\rho_c\simeq \sqrt{\frac{lR_s}{2}(1+\frac{3}{2}\sqrt{\frac{R_s}{2l}}+
\frac{3}{2}\frac{R_s}{2l}+...)}
\ee

Note that $\rho_c= R_s$ is what one  gets in flat 4 dimensions, whereas
$\rho_c= \sqrt{2G_5\mu}=\sqrt{R_s l/2}$ is what one gets in flat 5 dimensions,
so the formula is correct to zero-th order. 

So the mass of the black hole is 
\bea
&&M_{bh}\simeq \frac{\sqrt{s}}{\sqrt{2}}(1+\frac{l^2}{R_s^2}+...)\;\;\;l\ll 
R_s\nonumber\\
&&M_{bh}\simeq \frac{\sqrt{s}}{2}\sqrt{\frac{l}{R_s}}(1+\frac{3}{4}\sqrt{
\frac{R_s}{2l}}+...) \;\;\; l\gg R_s
\eea

Notice that the limit of small l is the limit in which the space is very 
4-dimensional (large exponential warping in the extra dimension), so the 
four-dimensional result should hold, and we find that (just small corrections
to the usual 4d result). The limit of large l is when the 
background space is approximately flat 5d space, so we have to modify 
the results to account for the creation of a 5d black hole. 
The condition $(\nabla \Phi)^2=4$ is independent of dimension, but 
it becomes $(\partial_i\Phi)^2+(\partial_y\Phi)^2=4$ in a general
dimension (with y being the transverse dimensions), and it will be modified 
for a general background. 

Thus in the general case the trapped 
surface is something in between two disks and 2 balls, so two fat disks, 
or flattened balls. In the 2 limiting cases, the trapped surface can be 
approximated by 2 disks or 2 balls, respectively. One can still  define 
the black hole projected onto 4 dimensions.

We will come back to the correct treatment in the next subsection, and we 
will see that whereas the zero-th order formulas are correct, the first 
order corrections get modified.

At nonzero b, applying the same approximation for finding $\rho_c$ as was used
in the flat case, the normal continuity 
 condition  $\partial_i \Phi_1\cdot \partial_i\Phi_2
=4$ becomes $\partial_i \Phi(\vec{x}-\vec{x}_1)\cdot \partial_i \Phi(\vec{x}-
\vec{x}_2)=4$, so we only get an extra factor of 
\be
\cos  \theta=1-\frac{b^2}{2\rho_c^2}
\ee
to the condition, so that now
\be
\frac{\rho_c}{R_s}=\sqrt{1-\frac{b^2}{2\rho_c^2}}
[1-\frac{2\rho_c/R_s}{\pi}\int_0^{\infty}
\frac{K_0'(y\rho_c/R_s)}{J_a^2(yl/R_s)+Y_1^2(y l/R_s)}]
\label{rhocb}
\ee
whereas the expression for the (very conservative) estimate of the trapped 
area, $S_{min}$, remains the same as a function of $\rho_c$ and b, 
\be
S_{min}= \sqrt{b^2\rho_c^2-\frac{b^4}{4}}
\ee
Expanding in the $l \ll R_s$ regime we get
\be
\frac{\rho_c^2}{R_s^2}=(1-\frac{b^2}{2\rho_c^2})[1+\frac{2 l^2}{\rho_c^2}
-\frac{l^4}{\rho_c^4}(8 ln \frac{\rho_c^2}{l^2}-13)+...]
\ee
so that 
\be
\frac{\rho_c^2}{R_s^2}=\frac{1}{2}(1+\sqrt{1-\frac{2b^2}{R_s^2}
+\frac{8l^2}{R_s^2}(1-\frac{b^2}{2R_s^2y_0})+
o(\frac{l^4}{R_s^4})})
\ee
(so that $b_{max}^2\simeq R_s^2/2(1+4l^2/R_s^2)$).
 
In the $l\gg R_s$ regime we have 
\be
\rho_c^2=\frac{lR_s}{2}\sqrt{1-\frac{b^2}{4\rho_c^2}}[1+\frac{3}{2}\frac{
\rho_c}{l}+\frac{3}{8}\frac{\rho_c^2}{l^2}+...]
\ee
The first term gives the equation
\be
x^3=a^2(x-\frac{b^2}{4});\;\;\;x=\rho_c^2;\;\;\;a=\frac{lR_s}{2}
\ee
Solving this equation and selecting the solution that gives $x=a$ in the 
limit of b=0, we get (also calculating the first two corrections)
\be
\frac{2\rho_c^2}{lR_s}=\frac{x}{a}= \alpha (1+\frac{3}{2}\sqrt{\frac{R_s\alpha}
{2l}}+\frac{3}{8}\frac{R_s\alpha}{2l}+...)
\ee
where 
\be
\alpha=\frac{1}{\sqrt{3}}(\Delta +\frac{1}{\Delta});\;\; \Delta= (-\beta
+\sqrt{-1+\beta^2})^{1/3};\;\; \beta=\frac{9b^2}{4\sqrt{3}lR_s}
\ee
If $\beta \leq 1$, then $\alpha$ is real and is
\be
\alpha=\frac{\cos \theta /3}{\sqrt{3}/2};\;\;{\rm where}\;\;\cos\theta =-\beta
\Rightarrow \Delta= e^{i\theta/3}
\ee
If $\beta>1$, the solution is complex, thus 
\be
b_{max}^2=\frac{4\sqrt{3}lR_s}{9}
\ee

\subsection{Correct treatment: generalizing the formalism \\
to curved higher dimensional background}

Let us try to understand what happens to the black hole area when we have 
a curved spacetime background of the RS type:
\be
ds^2=e^{-2|y|/l}[-dudv +dx_i^2]+dy^2
\ee
Let us denote $e^{-2|y|/l}=A$ and $g_{ij}=A\bar{g}_{ij}$ represents the 
metric in both x and y coordinates (transverse). Then a straightforward 
calculation along the lines of the flat space case finds the vector 
normal to the surface is 
\be
\xi=-\frac{1}{4}\bar{g}^{ij}\partial_i\Psi\partial_j\Psi du 
-dv -\partial_i \Psi dx^i
\ee
and so similarly to the flat case the continuity condition for the normal 
is 
\be
\bar{g}^{ij}\partial_i\Psi\partial_j\Psi =4 \Rightarrow
(\nabla \Psi)^2+A(\partial_y\Psi)^2=4
\ee
(the relation fixing the boundary of the trapped surface, or its radius)

For these and the next relations it is necessary to calculate the coordinate 
transformation from the coordinate system
\be
ds^2=e^{-2|\bar{y}|/l}(-d\bar{u}d\bar{v}+d\vec{x}^2+h\delta(\bar{u})
d\bar{u}^2)+d\bar{y}^2
\ee
to the coordinate system without delta function discontinuities, up to 
order u (near u=0). The calculation is a straightforward but tedious 
generalization of the flat space case, and one finds after the coordinate 
transformation
\bea
&& \bar{u}= u\nonumber\\
&&\bar{v}= v + h\theta (u) +\frac{u\theta(u)}{4} (\partial_i h \partial_jh 
\bar{g}^{ij}+A (\partial_y h)^2)\nonumber\\
&& \bar{x}^i = x^i +\frac{u\theta(u)}{2}\bar{g}^{ij}\partial_j h \nonumber\\
&&\bar{y}=y +\frac{u\theta(u)}{2}A\partial_y h
\eea
that
\bea
&&ds^2=A[-dudv +dx_i^2+u\theta(u)\partial_i\partial_j h dx^i dx^j]
\nonumber\\&&+dy^2[1+
u\theta(u)A\partial_y^2h]+dydx^i u\theta (u) A \partial_i\partial_y h +
dydA u\theta (u) \partial_y h +o(u^2)
\eea
where
\be
A= e^{-2|\bar{y}|/l}+o(u)^2\Rightarrow 
A|_{u=0}=e^{-2|y|/l};\;\;\;dA|_{u=0}=-\frac{2}{l}A[dy+\frac{A}{2} \partial_y h 
du]
\ee

The convergence of the normals $\theta=g^{ij}D_i\xi_j$ is now again
\be
\theta=-\nabla^2(\Psi-h)
\ee
where $h_{uu}=h\delta(u)(\equiv\Phi \delta (u))$ and 
\be
\nabla^2=\frac{1}{A}\nabla^2_x+\partial_y^2-\frac{d}{l}sgn(y)\partial_y
\ee
Therefore we write 
\be
\Psi=\Phi+\zeta ;\;\;\;\nabla^2\zeta=0
\ee

So now the trapped surface is a surface $f(\rho, y)=0$ defined by both 
$\Psi=C$ (const) and by $\bar{g}^{ij}\partial_i\Psi\partial_j\Psi =4 $.
In the flat case the first implied $\rho=\rho_0$ and the second $\rho_0=R_s$.
But we also saw that the nonzero b case had the same problem as we have now:
find a surface C and a function $\zeta$ 
that satisfies both $\Psi=const.$ and $\nabla^2\Psi=4$ with $\Psi=\Phi+\zeta$. 

In general it is a hard problem, but at least perturbatively, in the two 
limits $l\rightarrow 0$ and $l\rightarrow \infty$ we expect to find approximate
disks and approximate balls, respectively (and fat disks in between). 
We would also expect that in the $l\rightarrow 0$ the surface is the 
same disk $\rho=R_s$ as for flat 4d space.

The formula for $\Phi$ (h) at nonzero y is (in \cite{emp}, it's 
not $\Phi$ but $\Phi e^{-2|y|/l}$), so
\be
\Phi=-R_s[log \frac{r^2}{l^2}-\frac{2l}{\pi}e^{2|y|/l}\int_0^{\infty}
dm K_0(mr)\frac{Y_1(ml)J_2(ml e^{|y|/l})-J_1(ml)Y_2(mle^{|y|/l})}
{J_1^2(ml)+Y_1^2(ml)}]
\ee
(and actually, this is defined up to a constant, so the $log \;r^2/l^2$ is 
conventional, we could have $log \;r^2/r_0^2$).

Then we have  
\bea
&&\partial_y \Phi|_{y=0}=
\frac{2R_s}{\pi}[-\frac{4}{l\pi}\int_0^{\infty} \frac{d(ml)}{ml}\frac{K_0(mr)}
{J_1^2(ml)+Y_1^2(ml)}\nonumber\\
&&+\frac{2}{l}\int_0^{\infty} d(ml)K_0(mr)
\frac{Y_1(ml)J_2(ml)-J_1(ml)Y_2(ml)}
{J_1^2(ml)+Y_1^2(ml)}]=0!
\eea
where we have used that $Y_{\nu}(x)J_{\nu+1}'(x)-J_{\nu}Y_{\nu +1}'(x)
=-2(\nu +1)/\pi x^2$, which we can easily deduce from the Bessel function 
properties. 

Then we find
\bea
&&\partial_y^2\Phi|_{y=0}= \frac{2R_s}{\pi^2}[-\frac{8}{l^2}\int_0^{\infty}
\frac{d(ml)}{ml}\frac{K_0(mr)}{J_1^2(ml)+Y_1^2(ml)}\nonumber\\
&&+\pi\partial_y^2\int_0
^{\infty}d(ml) K_0(mr) 
\frac{Y_1(ml)J_2(mle^{|y|/l})-J_1(ml)Y_2(mle^{|y|/l})}{J_1^2(ml)
+Y_1^2(ml)}]|_{y=0}
\eea

Let us now analyze the perturbation in $l/r$ (the space is approximately 
flat 4d)

Using the relation
\be
Y_{\nu}(x)J_{\nu +1}''(x)-J_{\nu}(x)Y_{\nu +1}''(x)=\frac{2}{\pi x}
(\frac{6}{x^2}-1)
\ee
which can be easily derived, and also the expansion
\be
J_1(x)\sim x/2;\;\;\;\pi Y_1(x)\sim -\frac{2}{x}+x log \frac{x}{2}+...
\ee
we find 
\be
\partial_y^2\Phi|_{y=0}=-\frac{4R_sl^2}{r^4}+o(l^4/r^4)
\ee

We also have 
\be
\Phi|_{y=0}=-2R_slog \frac{r}{l}+R_s\frac{l^2}{r^2}+...
\ee

Let us expand $\zeta $ near y=0 as 
\be
\zeta=\zeta_0(r)+\zeta_1(r)y+\frac{y^2}{2}\zeta_2(r)
\ee
Then at y=0 $\nabla^2\zeta=0$ implies
\be
\partial_x^2 \zeta_0(r)+\zeta_2(r)-\frac{d}{l}\zeta_1(r)=0
\ee
and we don't want to upset the flat space solution, so we will take 
$\zeta_0=0$ (otherwise the continuity condition $(\nabla\Psi)^2=4$
implies a different radius for the trapped disks). So $\zeta_2=\frac{d}{l}
\zeta_1$. 

From
\be
(\partial_i\Psi)^2+e^{-2|y|/l}(\partial_y\Psi)^2=4
\ee
we see that if $\partial_y\Psi$ has a y-independent piece, we will change
the continuity equation at y=0, and we don't want that to happen 
to leading order in l. As $\partial\Phi|_{y=0}=0$ already, we must put 
$\zeta_1=0$ to leading order, so at least $\zeta_1\sim o(l)$, 
which implies $\zeta_2=o(1)$ as well.  

Then
\be
\Psi=f+a y +\frac{y^2}{2}g+...
\ee
where
\bea
&&f=\Phi_{|y=0}=-2R_s log r/l +R_sl^2/r^2+..., \;\;\;a=\zeta_1 
\nonumber\\&&
g=\partial_y^2\Phi|_{y=0}+\frac{d}{l}\zeta_1=
-\frac{4R_sl^2}{r^4}+o(l^4/r^4) +\frac{d}{l} \zeta_1\equiv g_0 +\frac{d}{l}
\zeta_1 + ...
\eea
We have to check now that the two surfaces in (r,y) defined by $\Psi=const.$
and the normal continuity are the same to first nontrivial order in y and l. 
\be
\Psi=C= f+ a y +\frac{y^2}{2}g+...
\ee
and the other 
\bea
&&C'=4= (f'+y a'+\frac{y^2}{2}g'+...)^2+(1-\frac{2y}{l}+2\frac{y^2}{l^2}+...)
(a+yg+...)^2\nonumber\\
&&=f'^2+a^2+y(2a'f'-2\frac{a^2}{l}+2ag)+...
\eea
if a is nonzero and 
\be
C'=4= f'^2 +y^2(f'g'+ g^2)+...
\ee
if a=0.
If a=0, we get to order $y^2$ (first nontrivial) for $\Psi=C$
\be
2R_s log \frac{r}{r_0}-R_s\frac{l^2}{r^2}+...= y^2(-\frac{2R_sl^2}{r^4}) 
\ee
(we have traded C for $r_0$) and for the continuity equation
\be
(4-f'^2-a^2=)
4-4\frac{R_s^2}{r^2}(1+2\frac{l^2}{r^2})-a^2 =y^2(g^2+f'g')=
y^2(-4\frac{8R_s^2l^2}{r^6})
\ee
Notice that at l=0 the l.h.s. of the two equations would be $2R_s \delta 
r/r_0$ and $8\delta r/R_s$ respectively, so with  $r_0=R_s$ (from y=0)
the two equations are not the same. So we have to put a nonzero $\zeta_1$.

Also note that since the constant C (and hence $r_0$) is an arbitrary constant,
at y=0 but l nonzero we don't need to have the same l dependence in the 
two equations, we can absorb the unwanted l dependence in the redefinition of 
$r_0$. The l dependence of the radius $r_{max}$ is deduced from the continuity 
equation (which doesn't have a free parameter). 

Also note that a priori one could check the values for $\Phi$ and its y 
derivatives by using the alternative solution for $\Phi$ in \cite{emp}. 
We have tried to use perturbation 
theory on the alternate form (integral of ratio of K function)
of $\Phi$, but as Emparan noted, it is much harder to do so.
In particular, one has to use the freedom to add an arbitrary  constant to h
(this is related to a rescaling of u and v).

If now we put $a=\zeta_1\neq 0$ (and so $g=g_0+\frac{4a}{l}$),
the first order in y is linear, and by 
requiring that at l=0 we get the same y dependence in both equations we 
get the condition
\be
a\simeq \frac{R_s}{4}(2a g_0+\frac{6a^2}{l}-\frac{4R_s}{r}a')
\ee
Thus if we put
\be
a=\frac{\alpha R_s l}{r^2}
\ee
at l=0 and $r=r_{max}=R_s$ and since $g_0\sim o(l^2)$ is negligible,
 we get $3/2 \alpha=-1$, or 
$\alpha =-2/3$.

Then at $y=0, l\neq 0$ the condition $\Psi=C$ is irrelevant as we said, since 
we can redefine the constant C. Then from the second (continuity) equation 
we get 
\bea
&&(\frac{2R_s}{r_{max}})^2(1+\frac{2l^2}{r^2}+\frac{\alpha^2 l^2}{4r^2}+...)
=4\nonumber\\
&& \Rightarrow r_{max}^2\equiv \rho_c^2=R_s^2 (1+\frac{19}{9}\frac{l^2}{r^2}
+...)\Rightarrow M_{bh}\simeq \frac{\sqrt{s}}{2}(1+\frac{19}{18}\frac{l^2}{
r^2}+...)
\eea
and in the treatment of the previous subsection we had thus neglected 
the $\alpha^2$ term, the equation needed to be modified, but the sign of the 
correction is the same. 

We can now also correct the calculation at nonzero b, by just putting the 
familiar $\cos \theta$ term
\be
\frac{\rho_c^2}{R_s^2}= (1-\frac{b^2}{2\rho_c^2})(1+\frac{l^2}{\rho_c^2}
(1+\frac{\alpha^2}{8})+...)
\ee
from which we get
\be
\frac{\rho_c^2}{R_s^2}= \frac{1}{2}(1+\sqrt{1-\frac{2b^2}{R_s^2}+\frac{8l^2}
{R_s^2}(1+\frac{\alpha^2}{8})(1-\frac{b^2}{2R_s^2y_0})+...})
\ee
The maximum impact parameter (and thus the scattering cross section
$\sigma= \pi b_{max}^2$) gets also
modified
\be
b_{max}^2\simeq \frac{R_s^2}{2}[1+\frac{4l^2}{R_s^2}(1+\frac{\alpha^2}{8})]
\ee 
The perturbation for $l\gg r$ (around flat 5d) will be left for future 
work.

\section{Aichelburg-Sexl solution in AdS background \\
and scattering analysis}

In this section we will analyze the case of an A-S wave in AdS (for future
application to the gauge-gravity duality). First, we have to derive the 
solution for the A-S wave inside AdS. 

\subsection{Aichelburg-Sexl solution in AdS bacground}

Let us notice that \cite{dh} analyzed putting A-S shockwaves in more 
general backgrounds, of the type
\be
ds^2=2A(u,v)du dv + g(u, v) h_{ij}(x^i) dx^i dx^j
\label{type}
\ee

The calculation of the A-S solution in this background, with 
a source=massless photon at $u=0, \rho=0$ was done as in 
flat background, just by gluing two regions at u=0 with a shift $\Delta v = f
=f(x^i)$. In \cite{dh}, it was  found that the Einstein equations are 
satisfied if 
\bea
&&A_{,v}=0=g_{,v}\nonumber\\
&&\frac{A}{g} \Delta f -\frac{g_{, uv}}{g} f = 32 \pi p G A^2 \delta (\rho)
\eea
Indeed, in Minkowski background (A=-1/2, g=1)
one finds the Aichelburg-Sexl solution, $\Delta f=-16\pi p G \delta^{(2)} 
(\rho)$.
Notice that if the equations are not satisfied, it just means that 
one can't find a solution for the ansatz taken. For example, spherical 
sourceless (p=0) waves of this type in flat space 
are excluded ($A=-1/2, g=r^2=(u-v)^2/4$ doesn't satisfy the conditions), 
but Penrose found another type of solution.

The authors of \cite{dh} 
were able to find such shockwaves in the Schwarzschild solution in 
Kruskal-Szekeres coordinates,
\bea
&&ds^2=-32 \frac{m^3}{r}e^{-r/2m}dudv +r^2(d\theta^2+\sin ^2 \theta d\phi^2)
\nonumber\\
&&uv \equiv -(r/2m-1)e^{r/2m}
\eea
namely
\be
f(\theta, \phi)= k \int_0^{\infty}\frac{\sqrt{1/2}cos (\sqrt{3}s/2)}{(\cosh s
-\cos \theta)^{1/2}}ds
\ee

Notice that if one would like to put AdS in the form in (\ref{type}), 
one can't:
For a  schockwave moving on the brane, the AdS background would be written as 
\be
ds^2=\frac{1}{z^2}(dudv+ d\vec{x}^2_2+dz^2)
\ee
which is not of the desired form, whereas for a wave moving in the z direction
\be
ds^2=\frac{dudv+d\vec{x}_3^2}{(u-v)^2}
\ee
which doesn't satisfy the conditions. But there could still be a solution of 
a different type.

Note that neither the previous metric nor the global AdS form 
\be
ds^2=l^2(-dt^2 cosh^2 \rho + d\rho^2 +sinh^2 \rho d\Omega_3^2)
\ee
nor the other forms
\be
ds^2= \frac{l^2}{cos^2\theta}(-dt^2+d\theta^2 +sin^2\theta d\Omega_3^2)
\ee
or (with $r/l= sinh \rho =\tan \theta$)
\be
ds^2= - d\tau^2 (1+\frac{r^2}{l^2})+\frac{dr^2}{1+\frac{r^2}{l^2}}+r^2 d
\Omega_3^2
\label{adsflat}
\ee
help us in putting AdS into the form desired by \cite{dh}, so we do need 
something else.

Indeed, we will see that instead 
we can follow closely the calculation of Emparan \cite{emp}, 
so we will describe it, modifying it for our purposes. 

Emparan \cite{emp} uses the metric of the one-brane RS model, perturbed 
with a general gravitational wave, in the form
\be
ds^2=e^{-2|y|/l}(-dudv +d\vec{x}^2 +h_{uu}(u, x^i, y)du^2)+dy^2
\label{rspert}
\ee
But all we have to do in order to go to AdS is to replace $|y|\rightarrow y$. 
Then, under the coordinate transformation
\be
y/l= ln \; z/l
\ee
we would get
\be
ds^2= \frac{l^2}{z^2}(-dudv +d\vec{x}^2 +h_{uu}(u, x^i, y)du^2+dz^2)
\ee
which is the form that we wanted to obtain using the \cite{dh} formalism.

But \cite{emp} gives the Einstein tensor for the RS metric (\ref{rspert}) as 
\bea
&& G_{yy}= \frac{d(d-1)}{2l^2} g_{yy}\nonumber\\
&& G_{\mu\nu}= (\frac{d(d-1)}{2l^2}-\frac{2(d-1)}{l}\delta(y))g_{\mu\nu}
\nonumber\\&&
-\frac{1}{2}\partial_{\mu}u\partial_{\nu}u [ e^{-2|y|/l}(\partial_y^2-
sgn(y)\frac{d}{l}\partial_y)+\nabla_x^2]h_{uu}
\eea
where we have actually corrected the \cite{emp} result by putting the 
sgn(y) function.  In the AdS case however, 
the sgn(y) is absent (since it came from $\partial_y|y|$).

The RS equations in the absence of $h_{uu}$ are 
\be
G_{AB}= \Lambda g_{AB}+ \lambda\delta(y) g_{\mu\nu}\delta_{AB}^{\mu\nu}
\ee
(cosmological constant $\Lambda$ in the bulk and on the brane $\lambda$
= brane tension)
and can be seen to be satisfied, we could read out what $\Lambda$ 
and $\lambda$ are. Then note that the equation for $h_{uu}$ is linear.

In our case,
adding the energy-momentum tensor of a photon of momentum p, (which 
will generate the A-S metric), travelling at fixed $x^i$ and fixed 
radial position in AdS, $y_0$, 
\be
t_{AB}=p\delta(u) \delta^{d-2}(x^i)\delta(y-y_0) \delta_{AB}^{uu}
\ee
we get an equation, with $h_{uu}\equiv \Phi \delta(u)$
\be
-\frac{1}{2} [ e^{-2y/l}(\partial_y^2-\frac{d}{l}\partial_y)+\nabla_x^2]
\Phi= 8\pi G_{d+1} p \delta^{d-2} (x^i)\delta(y-y_0)
\ee
Note that the flat space limit $l\rightarrow \infty$ gives the correct 
result, $-1/2\partial_i^2 h=8\pi G_{d+1} p \delta^{d-1}(x)$.

Going to 4d Fourier space
\be
\Phi(q,y)= \int d^{d-2}x e^{-iq\cdot x} \Phi(x,y)
\ee
and similarly for $t_{uu}$, one obtains 
\be
\Phi(q,y)''-\frac{d}{l}\Phi(q,y)'-q^2e^{2y/l}\Phi(q,y)=-16\pi p G_{d+1}
\delta (y-y_0)
\label{hsol}
\ee

Going back to Emparan's case \cite{emp}, the previous equation
 would have $ d/l \; sgn(y)$ and 
$e^{2|y|/l}$. The solution to that equation in Emparan's case is
\be
A e^{\frac{d|y|}{2l}}K_{d/2}(e^{|y|/l}lq)
\ee
where the Bessel function K was chosen among the 2 solutions to the Bessel 
equation because of the boundary conditions: one wanted that at $y\rightarrow
\infty$ the solution dies off, not blows up ($I_{d/2}$, the other solution, 
blows up exponentially at infinity). The $|y|$ in $e^{d|y|/2l}$ was because of 
the $sgn(y)$ in the equation, and the $|y|$ 
in the $e^{|y|/l}$ argument was due 
to the $e^{2|y|/l}$ in the equation. Then both at $y=\infty$ and $-\infty$ 
we need the behaviour of $K_{\nu}(x)$ for $x\rightarrow\infty$.

Finally, the constant is fixed by normalizing the coefficient of the 
delta function
\be
A(\frac{d}{2l}K_{d/2}(lq)+qK'_{d/2}(lq))=-8\pi G_{d+1}p
\ee
and using an identity for Bessel functions A can be put to a simpler form. 
Also using a more general energy momentum tensor for the momentum space wave,
$t_{uu}(q)\delta(y)$ one has 
\be
h_{uu}(q,y)= 8\pi G t_{uu}(q)e^{\frac{d|y|}{2l}}\frac{K_{d/2}(e^{|y|/l}lq)}
{qK_{d/2-1}(lq)}
\ee
For the photon energy momentum tensor, going back to x space and making 
the angular integrations, using
\be
\int d\Omega_{d-3}e^{iqr\cos \theta}= \Omega_{d-4}\int _0^{\pi} d\theta 
\sin^{d-4}\theta d\theta  e^{iqr\cos \theta}
=(2\pi)^{\frac{d-2}{2}}\frac{J_{\frac{d-4}{2}}(qr)}{(qr)^{\frac{d-4}{2}}}
\ee
one gets 
\be
h_{uu}(u,r,y)= \frac{4G_{d+1}}{(2\pi)^{\frac{d-4}{2}}}p \delta(u)
\frac{e^{\frac{d|y|}{2l}}}{r^{\frac{d-4}{2}}}\int_0^{\infty}dq 
q^{\frac{d-4}{2}}J_{\frac{d-4}{2}}(qr)\frac{K_{d/2}(e^{|y|/l}lq)}{K_{d/2-1}
(lq)}
\ee

In our case, the generalization is very simple. There are no $|y|$ in the 
equation (\ref{hsol}),
so none in the solution. Again the solution at $y\rightarrow \infty$
has to decay, so we choose the Bessel function K for $y>y_0$. But now for 
$y_0>y\rightarrow -\infty$ we get the exponent of the Bessel function becoming
$K(x)$, $x\rightarrow 0$, for which $K_{\nu}(x)$ blows up as $x^{-\nu}$. 
Instead, the Bessel function $I_{\nu}(x)$ behaves smoothly, as $x^{\nu}$. 
So the solution for $y<y_0$ is with $I_{d/2}$ instead of $K_{d/2}$. The 
normalization of the delta function is also different. 

The solution is now of the type
\bea
\Phi&=& A_1 e^{\frac{dy}{2l}}K_{d/2}(e^{y/l}lq)\;\;\;y>y_0\nonumber\\
&=& A_2 e^{\frac{dy}{2l}}I_{d/2}(e^{y/l}lq)\;\;\;y<y_0
\eea
Continuity at $y_0$ gives
\be
\frac{A_1}{A_2}= \frac{I_{d/2}(e^{y_0/l}lq)}{K_{d/2}(e^{y_0/l}lq)}
\ee
and the jump in the derivative gives the delta function normalization
($(\Delta\Phi'(y_0))=\\-16\pi G_{d+1}p e^{2y_0/l}$). Using 
the Bessel function relations
\bea
&&zI_{\nu}'(z)+\nu I_{\nu}(z)=zI_{\nu-1}(z)\nonumber\\
&& zK_{\nu}'(z)+\nu K_{\nu}(z)=-zK_{\nu-1}(z)\nonumber\\
&& I_{\nu}(z)K_{\nu +1}(z)+I_{\nu +1}(z)K_{\nu}(z) =\frac{1}{z}
\eea
we finally get
\bea
h_{uu}(q,y)&=& 8\pi G_{d+1} t_{uu}(q)e^{\frac{dy}{2l}}e^{\frac{4-d}{2l}y_0}
K_{d/2}(e^{y/l}lq)
2lI_{d/2}(e^{y_0/l}lq)\;\;y>y_0\nonumber\\
&=&8\pi G_{d+1} t_{uu}(q)e^{\frac{dy}{2l}}e^{\frac{4-d}{2l}y_0}
I_{d/2}(e^{y/l}lq)2lK_{d/2}(e^{y_0/l}lq)\;\;y<y_0
\eea
So now going in x space, taking the usual photon energy-momentum tensor 
and making the angular integrations we get 
\bea
h_{uu}(u,r,y)&=& \frac{8G_{d+1}l}{(2\pi)^{\frac{d-4}{2}}}p \delta(u)
\frac{e^{\frac{dy}{2l}}e^{\frac{4-d}{2l}y_0}
}{r^{\frac{d-4}{2}}}\int_0^{\infty}dq 
q^{\frac{d-2}{2}}J_{\frac{d-4}{2}}(qr) K_{d/2}(e^{y/l}lq)I_{d/2}(e^{y_0/l}lq)
\;\;\;y>y_0\nonumber\\
&=&\frac{8G_{d+1}l}{(2\pi)^{\frac{d-4}{2}}}p \delta(u)
\frac{e^{\frac{dy}{2l}}e^{\frac{4-d}{2l}y_0}
}{r^{\frac{d-4}{2}}}\int_0^{\infty}dq 
q^{\frac{d-2}{2}}J_{\frac{d-4}{2}}(qr) I_{d/2}(e^{y/l}lq)K_{d/2}(e^{y_0/l}lq)
\;\;\;y<y_0\nonumber\\&&
\eea
Again, the last integration cannot be done, except on a certain hypersurface.
Indeed, we have the relation 
\be
\int _0^\infty dx x^{\nu+1}K_{\mu}(ax)I_{\mu}(bx) J_{\nu}(cx)= 
\frac{(ab)^{-\nu-1}c^{\nu} e^{-(\nu +1/2)\pi i}Q_{\mu-1/2}^{\nu +1/2}(\mu)}
{\sqrt{2\pi}(\mu^2-1)^{\nu /2+1/4}}
\ee
(where $Q_{\mu}^{\nu}(z)$ is the associated Legendre function of the second 
kind),
that is of the desired form, which is however valid only if $Re(a)>|Re(b)|+
|Im(c)|$, $Re(\nu)> -1$, $Re(\mu+\nu)>-1$ (all satisfied) and $2ab\mu = 
a^2+b^2+c^2$, which imposes a constraint. 

Thus we obtain
\be
h_{uu}(u,r,y)= C\frac{8G_{d+1}l}{(2\pi)^{\frac{d-4}{2}}}p \delta(u)
e^{\frac{y-y_0}{l}}l^{2-d}e^{\frac{4-d}{l}y_0}
;\;\;\;
C=\frac{i^{\frac{3-d}{2}}Q_{\frac{d-1}{2}}^{\frac{d-3}{2}}(\frac{d}{2})}
{\sqrt{2\pi}(\frac{d^2}{4}-1)^{\frac{d-3}{4}}}
\ee
(for both $y<y_0$ and $y>y_0$!) on the hypersurface
\be
r^2=l^2 e^{2y_0/l}(de^{\frac{y-y_0}{l}}-1-e^{\frac{2(y-y_0)}{l}})
\ee

One could presumably check this by the Aichelburg-Sexl procedure, namely 
of boosting the AdS black hole and then taking the limit where the mass of 
the black hole goes to zero as the boost goes to infinity. It is however 
quite difficult in practice.

\subsection{Scattering analysis}

Let us look now at the AdS scattering. Let us first obtain the limits of 
AdS-A-S wave. Defining as before $h_{uu}=\Phi\delta(u)$ we get 
\be
\Phi= \bar{C}\frac{e^{\frac{dy}{2l}}}{r^{d-2}}e^{\frac{4-d}{2l}y_0}
\int _0^{\infty}dz z^{\frac{d}{2}-1}
J_{\frac{d-4}{2}}(z)K_{d/2}(e^{y/l}\frac{lz}{r})I_{d/2}(e^{y_0/l}\frac{lz}{r})
\ee
with $\bar{C}=8G_{d+1}lp/(2\pi)^{\frac{D-4}{2}}$. As we can see, for $r\gg l$ 
the integral is dominated by the region of small argument of I and K and 
we can use 
\be
I_{\nu}(x)\sim (\frac{x}{2})^{\nu}\frac{1}{\Gamma (\nu +1)};\;\;\;
K_{\nu}(x)\sim \frac{\pi}{2 sin \nu \pi }\frac{(x/2)^{-\nu}}{\Gamma (-\nu +1)}
\Rightarrow K_n (x)\sim \frac{1}{2}(n-1)!(\frac{x}{2})^{-n}
\ee
But 
\be
\int_0^{\infty} dx x^{2n+1} J_0(x)=0
\ee
so we need to expand $I_2(bx)K_2(ax)$ up to the first term that is not of 
$x^{2n}$ type. We find 
\be
I_2(bx)K_2(ax)= \frac{b^2}{4a^2}+ct. x^2 +ct. x^4- \frac{1}{64}a^2 b^2 x^4 
log (x)+o(x^5)
\ee 
and using
\be
\int_0^{\infty} dx x^5 log(x) J_0(x)=-64
\ee
we get 
\be
\Phi= \frac{\bar{C}l^4}{r^6} e^{\frac{2}{l}(2y+y_0)}
\ee
Instead, when $r\ll l$ (actually, for $e^{y/l} l/r\gg 1$), 
we can use the large argument expansion of I and K,
\be
I_{\nu}(x)\sim \frac{e^x}{\sqrt{\pi 2x}};\;\;\;K_{\nu}(x)\sim \sqrt{\frac{
\pi}{2x}}e^{-x}
\ee
and obtain (for d=4)
\be
\Phi\simeq
 \frac{\bar{C}e^{\frac{3y-y_0}{2l}}}{2l}\frac{1}{\sqrt{r^2+l^2(e^{y/l}
-e^{y_0/l} )^2}}
\label{largel}
\ee
and so if we also have $y/l, y_0/l\ll 1$ we obtain as expected the 5d 
result
\be
\Phi\simeq
 \frac{C}{2l\sqrt{r^2+(y-y_0)^2}}
\ee
Note that the result in (\ref{largel}) can be obtained also if $r/l\sim 1,
e^{y/l}\gg 1$, which means $y/l\sim$ a few (not too large).

Another particular case of interest is $y=y_0$. Then we can do the integral 
at all values of r and obtain ($\bar{C}= 2 R_s l^2$, $R_s$ is 4d 
the Schwarzschild radius) 
\be
\Phi=2R_s[-1+\frac{r^2}{2l^2}e^{-2y_0/l}(-1+\sqrt{1+\frac{4l^2}{r^2
e^{-2y_0/l}}})+\frac{l^2}{r^2e^{-2y_0/l}}\frac{1}{\sqrt{1+\frac{4l^2}{r^2
e^{-2y_0/l}}}}]
\ee
and we can check that for $r\gg l$ (and $y_0/l\sim 1$ or $\ll 1$) we get 
\be
\Phi\simeq \frac{2R_s l^6}{r^6e^{-6y_0/l}}
\ee
same as the result that we obtain in this limit from the above answer
for all y ($\neq y_0$). 

We can also check that at $e^{y_0/l}l/r\gg 1$ we have 
\be
\Phi \simeq l R_s\frac{e^{y_0/l}}{r}
\ee
as we obtained from the formula at arbitrary y. 

Finally, let us now look at 't Hooft scattering in $AdS_5$ in the two limits. 
For $r\ll l$ or $l/r\sim 1, e^{y/l}\gg 1$ (so that $lq \gg 1$ or $lq \sim
1, e^{y/l}\gg 1$)
\be
\Phi
\simeq \frac{\bar{C}}{2l}
\frac{e^{\frac{3y-y_0}{2l}}}{\sqrt{r^2+l^2(e^{y/l}-e^{y_0/l})^2}
}
\ee
and hence (since $\delta= p_-^{(1)}\Phi$, and going to $z=qb\equiv qr$
variables and using $p_{-}^{(1)} p^{(2)}=s/4$)
\be
\delta (b,s)=\frac{G_{5}se^{\frac{3y-y_0}{2l}}q}{
\sqrt{z^2+l^2q^2(e^{y/l}-e^{y_0/l})^2}}
\ee
and thus if $\delta$ is small the amplitude is 
\bea
&&{\cal A} \simeq \frac{A G_{5}se^{\frac{3y-y_0}{2l}}}{
q}\int_0^{\infty}dz z \frac{1}
{\sqrt{z^2+l^2q^2(e^{y/l}-e^{y_0/l})^2}}J_{0}(z)\nonumber\\
&&= \frac{G_5}{2\pi} \frac{s}{\sqrt{t}}
e^{\frac{3y-y_0}{2l}}exp[-(\sqrt{t}l(e^{y/l}-e^{y_0/l}))]
\eea
where the exponent is therefore large. 

However, $\delta \ll 1$ means either $y\neq y_0$ and $G_4 s e^{3(y-y_0)/(2l)}
\ll 1$ or $y\simeq y_0$ and $G_4 s \frac{l}{r} e^{y_0/l}\ll 1$, so the only
possibility is $y\neq y_0<l $, $r\ll l$, $G_4 s\ll 1$, but we still want 
$G_4 s\sim 1$, but $<1$ for 't Hooft scattering, so it's not clear that 
there is a good regime in between.  

For $r\gg l$ (or rather $lq\ll 1$), we obtain in D=4
\be
\delta (b,s)\simeq 2\frac{G_5s l^5 e^{2(2y+y_0)/l}q^6}{z^6}
\ee
and therefore now $\delta$ is always small, so  
\be
{\cal A}\simeq \frac{A}{q^{D-2}}
\int_0^{\infty} dz z^{D/2-1}J_{D/2-2}(z)\delta (z)
\ee
Unfortunately, the result in D=4 is infinite, and to obtain the finite 
t-dependent piece we would need to get $\delta$ at general D, which seems to 
be quite difficult to do, but we can say that the result in $D=4+2\epsilon$ 
will change $(q/z)^6\Rightarrow (q/z)^{6+m\epsilon}$ and so 
\be
{\cal A}(D=4)= \frac{G_4 s l^6 t^2}{2\pi 2^7}e^{\frac{2}{l}(2y+y_0)}
\frac{m-2}{3-m} ln \; t \propto G_4 s l^6 t^2 ln \; t\;
e^{\frac{2}{l}(2y+y_0)} 
\ee

\section{Conclusions}

In this paper we have reanalyzed the question of black hole formation in 
the high energy collision of two particles via the classical scattering 
of two shockwaves. 

We have found that string corrections increase the horizon area. For the 
effective shockwave metric in \cite{acv93}, we have found that if we scatter 
head-on (at b=0) two such waves, each characterized by an impact parameter 
$b>R_s$, we obtain trapped surfaces which are deformed disks of area higher 
than the area obtained from A-S wave scattering. For the effective shockwave 
metric in \cite{ac}, in the case of of $R_S^2/Y\gg 1$ ($Y= \alpha ' 
log (\alpha ' s)$), we get an increase of the area of the black hole 
formed, as well as of the classical scattering cross section, $\sigma=
\pi b_{max}^2$,
while in the $R_s^2/Y\ll 1$ we get that the area of the formed 
black hole is of the order of Y (modified string scale), not $R_s^2$, 
so much larger. 

For higher dimensions, we have found a conservative 
approximation scheme for the area of the horizon formed which gives us a 
maximum impact parameter (indicative of the scattering cross-section, 
as we expect that $\sigma= \pi b_{max}^2$). We have thus obtained that 
 in D=4, $b_{max}= R_s/\sqrt{2}$, and in D=5 for instance 
$b_{max}\simeq 0.9523
R_s$, which is again a more conservative estimate as the one in \cite{eg}. 

What was more surprising was the fact that although graviton-graviton 
scattering should be described by the collision of two ideal sourceless
waves, given in the Khan-Penrose solution, there doesn't seem to be a 
horizon forming even at zero impact parameter. There is a theorem that 
a singularity will form in the future of any sourceless wave collision, 
yet we can't find a trapped surface, namely the usual trapped surface 
calculation doesn't have a solution. 
 We have speculated that maybe the gravitons cannot 
be described by sourceless waves at all, or maybe trapped surfaces 
are inherently different from the \cite{eg} case, namely that the 
surfaces form only in the interacting region $u>0,v>0$, not at the 
border (u=0, v=0) as in the photon scattering case.

We have extended the formalism to curved backgrounds. For more realistic 
scenarios, involving possible creation of black hole at accelerators 
for low fundamental scale, we have chosen the one brane Randall-Sundrum 
case. In the case that the 5th direction is highly curved, we have obtained 
just corrections to the flat 4d case, whereas for a weakly curved 5th 
direction, we have corrections about the 5d flat space black hole creation. 

Finally, we have found a solution for an Aichelburg-Sexl wave inside 
an AdS background, and we have calculated the scattering amplitude for 
't Hooft scattering in such a wave, at small and large distances r. 
This was done for later use \cite{inprog} for analysis of the gravity 
dual of QCD high energy scattering. 

{\bf Acknowledgements} We would like to acknowledge useful discussions 
with Matt Strassler, Radu Roiban and Antal Jevicki.
This research was  supported in part by DOE
grant DE-FE0291ER40688-Task A.

\end{document}